\documentclass[prd,twocolumn,superscriptaddress,floatfix,showpacs,preprintnumbers,amsmath,amssymb]{revtex4}
\usepackage[dvips]{graphicx}
\usepackage{epsfig}             % Allows inclusion of eps files.
\usepackage{dcolumn}            % used for alignment along decimal places in tables
\usepackage{bm}

\newcommand{\beq}[1]{\begin{equation} #1 \end{equation}}
\newcommand{\beqa}[1]{\begin{eqnarray} #1 \end{eqnarray}}

\newcommand{\pderiv}[2]{\frac{ \partial #1 }{ \partial #2 }}
\newcommand{\bfbar}[1]{\mathbf{\bar{#1}}}
\newcommand{\qbar}{\bfbar{q}}
\newcommand{\ibanez}{J.~$\mathrm{M}^{\underline{a}}$~Ib\'{a}\~{n}ez}
\newcommand{\marti}{J.~$\mathrm{M}^{\underline{a}}$~Mart\'{i}}

\newcommand{\aap}{{Astron.~Astrophys.}}                % Astronomy and Astrophysics
\newcommand{\physrep}{{Phys.~Rep.}}   % Physics Reports
\newcommand{\aaps}{{Astron.~Astrophys.~Supp.}}              % Astronomy and Astrophysics, Supplement
\newcommand{\cmp}{{Comm.~Math.~Phys.}}  %Communications in Mathematical Physics
\newcommand{\cqg}{{Class.~Quant.~Grav.}}  %Classical and Quantum Gravity 
\newcommand{\jcompphys}{{J.~Comp.~Phys}}  % Journal of Computational Physics

\newcommand{\grqc}[1]{{arxiv:gr-qc/#1}}
\newcommand{\lr}{{Living~Reviews}}
\newcommand{\physlettb}{{Phys.~Lett.~B}}
\newcommand{\pr}{{Phys.~Rev.}}                   % old Physical Review
\newcommand{\ijnmf}{{Int.~J.~Numer.~Methods~Fluids}}   % International Journal of Numerical Methods for Fluids

\begin{document}

\title{Type II critical phenomena of neutron star collapse}

\preprint{AEI-2007-140}

%\date{\today}

\author{Scott C. Noble}
\email{scn@jhu.edu}
\homepage{http://www.pha.jhu.edu/~scn/}
\affiliation{Department of Physics and Astronomy, Johns Hopkins University, 
3400 North Charles Street, Baltimore, MD 21218, USA}

\author{Matthew W.\ Choptuik}
	\email{choptuik@physics.ubc.ca}
   \homepage{http://bh0.physics.ubc.ca/~matt/}
	\affiliation{
		Dept.~of Physics and Astronomy, University of British Columbia, Vancouver BC, V6T 1Z1 Canada
	}
	\affiliation{
		CIFAR Cosmology and Gravity Program
	}
	\affiliation{
		Max-Planck-Institut f\"ur Gravitationsphysic, Albert-Einstein-Institut, Am M\"uhlenberg 1, D-14476 Golm, Germany
	}

%%%%%%%%%%%%%%%%%%%%%%%%%%%%%%%%%%%%%%%%%%%%%%%%%%%%%%%%%%%%%%%%%%%%%%%%%%%%%%%%%%%%%
%% ABSTRACT:   %%%%%%%%%%%%%%%%%%%%%%%%%%%%%%%%%%%%%%%%%%%%%%%%%%%%%%%%%%%%%%%
%%%%%%%%%%%%%%%%%%%%%%%%%%%%%%%%%%%%%%%%%%%%%%%%%%%%%%%%%%%%%%%%%%%%%%%%%%%%%%%%%%%%%

\begin{abstract}
We investigate spherically-symmetric, general 
relativistic systems of collapsing perfect fluid distributions.
We consider neutron star models 
that are driven to collapse by the addition of an initially 
"in-going" velocity profile to the nominally static star solution.  
The neutron star models we use are Tolman-Oppenheimer-Volkoff 
solutions with an initially isentropic, gamma-law equation of state.
The initial values of 1) the amplitude of the velocity profile, and 2) the 
central density of the star, span a parameter space, and we focus only on that 
region that gives rise to Type~II critical 
behavior, wherein black holes of arbitrarily small mass can be formed.  
In contrast to previously published work, 
we find that---for a specific value of the adiabatic index ($\Gamma = 2$)---the observed 
Type~II critical solution 
has approximately the same scaling exponent as that calculated for an
ultrarelativistic fluid of the same index.
Further, we find that the 
critical solution computed using the ideal-gas equations of state 
asymptotes to the ultrarelativistic critical solution. 
\end{abstract}

%Any other classifications??

\maketitle

%%%%%%%%%%%%%%%%%%%%%%%%%%%%%%%%%%%%%%%%%%%%%%%%%%%%%%%%%%%%%%%%%%%%%%%%%%%%%%%%%%%%%
%% CHAPTER:   %%%%%%%%%%%%%%%%%%%%%%%%%%%%%%%%%%%%%%%%%%%%%%%%%%%%%%%%%%%%%%%
%%%%%%%%%%%%%%%%%%%%%%%%%%%%%%%%%%%%%%%%%%%%%%%%%%%%%%%%%%%%%%%%%%%%%%%%%%%%%%%%%%%%%
\section{Introduction}
\label{sec:introduction}

%TYPE-II
Critical phenomena in general relativity involves the study of solutions---called \emph{critical} 
solutions---that lie at the boundary between 
black hole-forming and black hole-lacking spacetimes.  
(See \cite{choptuik-1998,gundlach,gundlach-rev2} for reviews.)
Published work in  general relativistic critical phenomena  
began  over a decade ago with a detailed numerical examination of the collapse 
dynamics of a massless scalar field,
minimally coupled to the general relativistic gravitational field~\cite{choptuik-1993}.
This first study in critical phenomena touched upon the three fundamental aspects of 
black-hole-threshold critical behavior: 1) universality and 2) scale invariance of the critical solution with 3) power-law 
behavior in its vicinity.   All three of these features
have now been seen in a multitude of matter models, such as 
perfect fluids \cite{evans-coleman,neilsen-crit,brady_etal}, an
$\mathrm{SU}(2)$ Yang-Mills model~\cite{choptuik-chmaj-bizon,choptuik-hirshmann-marsa}, 
and collisionless matter~\cite{rein-etal-1998,olabarrieta-choptuik} to name a few.
It was eventually found that 
there are two related yet distinct types of critical phenomena: Type~I and Type~II, so named because of the 
similarities between critical phenomena in general relativity and those of statistical mechanics.  

Type~II behavior was the first to be discovered \cite{choptuik-1993}, and entails critical solutions that are 
either continuously self-similar (CSS) or discretely self-similar (DSS).  
Super-critical solutions---those that form black holes---give rise to black holes with 
masses that scale as a power-law, 
\beq{
M_\mathrm{BH} \propto \left|p - p^\star\right|^\gamma  \, ,
\label{mass-scaling}
}
implying that arbitrarily small black holes can be formed.  Here, $p$ parameterizes a $1$-parameter family 
of initial data with which one can tune toward the critical solution, 
located at $p=p^\star$, and $\gamma$ is the scaling exponent of the critical behavior.  
Since $M_\mathrm{BH}(p)\to 0$ as 
$p\to p^\star$, 
this type of critical behavior was named ``Type~II'' since it parallels 
Type~II (continuous) phase transitions of statistical mechanics.  

As in the statistical mechanical case, 
there is also a Type~I behavior in gravitational collapse, 
where the black hole mass ``turns on'' at a finite value.  As might be expected,
Type~I critical solutions are quite different from their Type~II counterparts, tending to be 
meta-stable star-like solutions that are static or periodic.  In this paper, attention
is restricted to Type~II behavior; results from our study of Type I transitions 
in our model are reported in a separate paper~\cite{noble-choptuik2}.

The accepted picture describing the scaling behavior seen in Type II critical collapse
was suggested by Evans and Coleman \cite{evans-coleman}, who
computed the critical solution for a radiation fluid (fluid 
pressure, $P$, and density, $\rho$, related by $P=\rho/3$) in two distinct ways.  
First, using a code that solved the full set of partial differential equations
(PDEs) for the fluid and gravitational field, and by tuning an initial data
parameter as sketched above, Evans and Coleman were able to compute a strong field 
solution that sat at the threshold of black hole formation, as well as establish 
a scaling law of the form~(\ref{mass-scaling}) .  Furthermore, the results 
of this numerical experiment provided compelling evidence that the threshold solution was 
continuously self-similar.  Second, by adopting the assumption of continuous 
self-similarity as an ansatz, Evans and Coleman reduced the set of PDEs governing their 
model to a set of ODEs, from which a precisely CSS solution was calculated.   The solutions 
computed using these two completely different techniques were found to agree 
extremely well.  Crucially, it was argued that the observed scaling behavior of 
the mass above threshold could be explained using linear perturbation theory about 
a background given by the CSS critical solution. 
Such an analysis was  carried out by Koike et al. \cite{koike-etal-1995} (for the 
radiation fluid), who showed that the scaling exponent, $\gamma$, was the inverse 
of the Lyapunov exponent of the critical solution's single, unstable eigenmode.  

Subsequent work showed that $\gamma$
was not a truly universal constant, but that its value could depend on the specifics 
of the matter model used.  The first evidence for this non-universality in 
scaling behavior was given in concurrent works by 
Maison \cite{maison-1996} and Hara et al. \cite{hara-etal-1996}.  Using similar 
methods to those of \cite{koike-etal-1995,koike-etal-1999}, they found that $\gamma$ for
an ``ultrarelativistic'' fluid with equation of state (EOS) 
\beq{
P=\left(\Gamma-1\right)\rho
\label{ultra-eos}}
is dependent on the adiabatic index, $\Gamma$. 

Most of these investigations, however, have involved ultrarelativistic fluids that are explicitly scale-free. 
The reason for the predominance of this type of fluid is due to the fact that  Cahill and Taub \cite{cahill-taub}
showed that only those perfect fluids which have state equations of the form of Eq.~(\ref{ultra-eos})---i.e.~the so-called ultrarelativistic EOS---can give rise to spacetimes that admit a 
homothetic symmetry (i.e.~which are self-similar).
Hence, it is not completely unreasonable to expect that Type~II, CSS critical solutions would 
only appear in such fluids, or at least in fluids that admit an ultrarelativistic limit.  
To study this conjecture, Neilsen and one of the current authors \cite{neilsen-crit} considered the evolution 
of a typical perfect fluid  with equation of state, 
\beq{
P=\left(\Gamma-1\right)\rho_\circ \epsilon
\label{ideal-eos}
}
that introduces a length scale into the field equations. Here, $P$ is the pressure, 
$\rho_\circ$ is the rest-mass energy density, and $\epsilon$ is 
the specific internal energy density.  It was argued in~\cite{neilsen-crit} that
Type~II critical collapse scenarios are 
typically kinetic energy dominated, entailing increasingly large central pressures that
maintain
the tenuous 
balance between the matter dispersing from the origin and collapsing to a black hole.  Therefore, if was reasoned 
that should one be able 
to give the fluid sufficient kinetic energy, then it would naturally enter 
an ultrarelativistic phase.  Specifically, if the fluid undergoes a collapse
such that $\epsilon \rightarrow \infty$ dynamically, then  $\rho_\circ$ will effectively become 
negligible in the equations of motion (EOM) and the system will be able to follow a scale-free---hence 
self-similar---evolution.  To see if this hypothesis was correct, compact distributions
of perfect fluid, with $P=0.4\rho_\circ \epsilon$ ($\Gamma=1.4$) were collapsed,
and the calculations were tuned to a 
threshold solution. 
The critical solution thus obtained by solving the full set of PDEs
closely 
matched the precisely self-similar solution, which was calculated by assuming that a model governed by 
the ultrarelativistic EOM had an exact homothetic symmetry.  Further, it was found that the scaling 
exponent, $\gamma$, 
defined by Eq.~(\ref{mass-scaling}) matched that of the ultrarelativistic critical solution with
 $\Gamma=1.4$. 
Since the ultrarelativistic fluid exhibited Type~II phenomena 
for all considered values of the adiabatic index in the range $1.05 \lesssim \Gamma \le 2$, 
the results of \cite{neilsen-crit} suggested that the 
Type II ideal-gas critical solution
for \emph{any} $\Gamma$ in that range
should be the same as that for an ultrarelativistic fluid with the same $\Gamma$.  

This hypothesis is not without precedence, since several models have been found to exhibit 
DSS or CSS collapse, even when explicit length scales are present.  For instance, 
one of us found Type~II behavior for the case of a 
collapsing \emph{massive} scalar field~\cite{choptuik-1994}---that 
is a scalar field with potential $V(\phi)=\frac{1}{2}m^2\phi^2$---even though 
the model has an explicit length 
scale set by $1/m$.  The heuristic argument presented in~\cite{choptuik-1994}
is that the potential term is 
naturally bounded since $\phi$ itself is bounded in the critical regime, 
but that the kinetic term---$\Box \phi$---diverges in the critical limit.
Hence, the kinetic term overwhelms the potential term and essentially makes the critical evolution scale-free.

The single study exhibiting Type~II behavior in perfect fluid collapse
with an ideal gas EOS~\cite{neilsen-crit}
remained unverified until work by Novak~\cite{novak}.  To determine the possible 
range in masses of nascent black holes formed from stellar collapse, Novak performed a 
parameter space survey using Tolman-Oppenheimer-Volkoff (TOV) 
solutions with $\Gamma=2$, varying the 
overall amplitude, $U_\mathrm{amp}$, of an otherwise fixed initial coordinate 
velocity profile for the fluid in order to generate critical solutions.
The Type~II behavior observed was quantified by fitting to the 
typical black hole mass scaling relation (\ref{mass-scaling}), 
where $p$ was identified with $U_\mathrm{amp}$.
Significantly, Novak was able to observe scaling behavior even with a 
realistic equation of state formulated by Pons et al. \cite{pons-etal-2000}.  
This was somewhat surprising, since there were some expectations that
Type~II phenomena would not be observed when using realistic equations
of state \cite{gundlach-rev2}. 
However, provided that the equation of state admits an ultrarelativistic limit,
as is apparently the case for Novak's calculations, the heuristic argument sketched above
suggests that
one {\em should} expect to 
see Type II transitions.

Although Novak observed Type~II behavior, he did not find the same scaling exponent as 
had been observed for the $\Gamma=2$ ultrarelativistic fluid in the study 
described in~\cite{neilsen-crit}.  In addition, 
he claimed that $\gamma$ was 
a function of 1) the central rest-mass density, $\rho_c$, which (as described in the next 
section)
parameterizes the initial star solution, and 2) 
the EOS used.  
He observed that the fit to Eq.~(\ref{mass-scaling}) worsened as $\rho_c$ increased
to that of the maximum mass solution, and that it eventually broke down completely.  Specifically, 
he found for the ideal-gas EOS (\ref{ideal-eos})
\beq{
\gamma \ \simeq \  0.52 \, , \label{novak-exponent1}
}
and when using the realistic EOS 
\beq{
\gamma \ \simeq \ 0.71   \, . \label{novak-exponent2}
}
These values are significantly different from the values most recently calculated 
with the $\Gamma=2$ ultrarelativistic fluid~\cite{brady_etal} using a variety of 
methods:
\beq{
\gamma \ \simeq \ 0.95 \pm 0.1  \, .
\label{brady-exponent}
}
Here we have taken the average of the three independent values calculated 
in~\cite{brady_etal} 
and the quoted uncertainty is the standard deviation of those values, and 
does not include the systematic errors inherent in the distinct calculations.

However, Novak clearly states in~\cite{novak} that his code was not designed to simulate the formation of very small black holes, 
and apparently
was only able to tune to a precision of $|p-p^\star|/p\simeq10^{-3}$.  In this paper, we 
reexamine
the Type~II behavior in this particular system in order to 
check the claims of~\cite{novak}, and obtain 
what we claim is an improved measurement of the scaling exponent.  
Different families of initial data are used to 
demonstrate the universality of our computed critical solution.  
Also, we compare the critical configuration calculated
from a near-threshold neutron star collapse to the critical solution obtained 
using an explicitly ultrarelativistic fluid. 
To more accurately study
CSS behavior as the black hole threshold is approached, we
employ mesh refinement techniques and non-uniform discretization.  
Further, we implement methods that 
improve the accuracy of the transformation from so called conservative variables 
to primitive variables that is required in our numerical analysis.  

The remainder of this paper is structured as follows.
Section~\ref{sec:theoretical-model} introduces the model and equations used to describe our collapsing neutron stars.  
In Sec.~\ref{sec:numerical-techniques}, we describe the numerical methods used to 
solve the coupled fluid and gravitational PDEs, and 
include a 
discussion of an instability we observe near the critical threshold.  
In Sec.~\ref{sec:type-ii-critical}, we analyze
the observed Type II behavior and compare it to previously published work. 
We then conclude in Sec.~\ref{sec:concl-future-work} with some closing remarks and 
notes on anticipated future work. 

Geometrized units, $G = c = 1$, are used throughout, and our tensor notation follows \cite{wald}. 
%%%%%%%%%%%%%%%%%%%%%%%%%%%%%%%%%%%%%%%%%%%%%%%%%%%%%%%%%%%%%%%%%%%%%%%%%%%%%%%%%%%%%
%%%%%%%%%%%%%%%%%%%%%%%%%%%%%%%%%%%%%%%%%%%%%%%%%%%%%%%%%%%%%%%%%%%%%%%%%%%%%%%%%%%%%
%%%%%%%%%%%%%%%%%%%%%%%%%%%%%%%%%%%%%%%%%%%%%%%%%%%%%%%%%%%%%%%%%%%%%%%%%%%%%%%%%%%%%
%% Section:   %%%%%%%%%%%%%%%%%%%%%%%%%%%%%%%%%%%%%%%%%%%%%%%%%%%%%%%%%%%%%%%
%%%%%%%%%%%%%%%%%%%%%%%%%%%%%%%%%%%%%%%%%%%%%%%%%%%%%%%%%%%%%%%%%%%%%%%%%%%%%%%%%%%%%
\section{Theoretical Model}
\label{sec:theoretical-model}

As in many previous critical phenomena studies in spherical symmetry  
\cite{brady_etal,choptuik-1993,choptuik-chmaj-bizon,neilsen-crit,novak}, we employ the so-called 
polar-areal metric
\beq{ 
ds^2 = - \alpha\left(r,t\right)^2 dt^2 + a\left(r,t\right)^2 dr^2 
+ r^2 d\Omega^2  \, , \label{metric}
}
where $\alpha$ is often referred to as the lapse function. 
We use the perfect fluid approximation for the matter model of our neutron stars, so the 
stress-energy tensor takes the form
\beq{
\hat{T}_{a b} = \left( \rho + P \right) u_a u_b + P g_{a b} \, .  \label{stress}
}
Here, $u^a(r,t)$ is the 4-velocity of a given fluid element, $P(r,t)$ is the isotropic pressure, 
$\rho(r,t) = \rho_\circ \left(1 + \epsilon\right)$ is
the energy density, $\rho_\circ(r,t)$ is the rest-mass energy density, and 
$\epsilon(r,t)$ is the specific internal energy. 

Modern computational methods that cast the hyperbolic fluid equations of motion 
in conservation form, and that use information concerning the characteristics of 
the equations, have been used very successfully in the modeling of 
highly-relativistic flows with strong gravity
(see  \cite{banyuls,font-review,font-etal2,neilsen,romero} 
for a small, but representative, selection of papers on this topic).  
Here we adopt the formulation used by
Romero et al. \cite{romero}, with a change of variables similar to that 
performed in~\cite{neilsen}. 
The formulation described in~\cite{romero} has been used extensively for 
treating relativistic flows in the presence of strong gravitation fields, and appears to work
quite well in such instances.

The EOM for the fluid are derived from the local conservation equations 
for energy and baryon number which 
are, respectively, 
\begin{equation}
\nabla_a {T^a}_b = 0  \, , \label{energy-cons}
\end{equation}
\begin{equation}
\nabla_a \left( \rho_{\circ} u^a \right) = 0  \, . \label{current-cons}
\end{equation}
Rather than working directly with the components of the fluid 4-velocity, it is more 
useful to employ
the radial component of the Eulerian---or physical---velocity 
of the fluid as measured by an Eulerian observer:
\beq{
v(r,t) = \frac{ a u^r }{\alpha u^t} \, ,
}
where $u^\mu = \left[ u^t, u^r, 0, 0\right]$.  
The associated ``Lorentz gamma function'' is defined by 
\beq{ 
W(r,t) = \alpha u^t  \, ,
\label{w}
}
and satisfies the usual relation
\beq{
W^2 = \frac{1}{1 - v^2} \,\, ,  \label{vwrelation}
}
since the 4-velocity is time-like and unit-normalized, i.e.~$u^\mu u_\mu=-1$.
Adopting a notation where bold face symbols denote state vectors, 
the fluid EOM can be cast in conservative form as
\beq{ 
\partial_t \mathbf{q} 
+ \frac{1}{r^2} \partial_r \left( r^2 X \mathbf{f} \right) = \bm{\psi} 
\, , \label{conservationeq}
}
where $\mathbf{q}\equiv \mathbf{q}(r,t)$ is a state vector of \emph{conserved} variables,  
and 
$\mathbf{f}\equiv\mathbf{f}(\mathbf{q})$ and $\bm{\psi}\equiv\bm{\psi}(\mathbf{q})$ 
are, respectively, the flux and source state vectors. 
Our choice for the
conserved variables is the one used almost exclusively in the field 
\cite{brady_etal,neilsen-crit,neilsen,novak,romero}, 
namely $\mathbf{q} = \left[ D(r,t), S(r,t), \tau(r,t) \right]$ with
\beqa{
D & = & a \rho_{\circ} W  \nonumber \, ,\\
S & = & \rho_\circ h W^2 v  \nonumber \, ,\\
\tau & = & E - D  \label{D-S-tau-E} \, ,\\ 
E(r,t)  & =  & \rho_\circ h W^2 - P  \, ,  \nonumber
}
and where $h(r,t) \equiv 1 + \epsilon + P/\rho_\circ$ is the specific 
enthalpy of the fluid.
$D$, $S$, $E$, and $\tau$ can be thought of as the rest-mass density, momentum density, 
total energy density, and internal energy density, respectively, as measured in 
a Eulerian-frame defined by the ADM 
slicing.  We found that for extremely relativistic 
flows near the threshold of black hole formation, this formulation was not very stable. 
We therefore use a different formulation 
motivated by~\cite{neilsen}, where it was found that evolving 
$\tau \pm S$ allowed for a more precise calculation since $\tau \sim S$ in the ultrarelativistic regime. 
We thus define new variables
\beq{
\Pi(r,t)  \ \equiv \  \tau + S  \, , \quad  \ \Phi(r,t)  \ \equiv  \ \tau - S 
\label{Pi-Phi-ideal}
}
and the state vectors become
\beqa{ 
\mathbf{q} =  \left[ \begin{array}{c} D \\ \Pi \\ \Phi \end{array}\right]  
& \ , \ &
\mathbf{f} = \left[ \begin{array}{c} D v \\ v \left( \Pi + P \right) + P
        \\ v \left( \Phi + P \right) - P \end{array} \right] \nonumber  \ , \ \\ 
\bm{\psi} = \left[ \begin{array}{c} 0 \\ \Sigma \\ -\Sigma \end{array} \right]
& \ , \ &
\mathbf{w} = \left[ \begin{array}{c} P \\ v \\ \rho_\circ \end{array} \right]  \, . 
\label{ideal-piphi-state-vectors}
}
The elements of the vector $\mathbf{w}$ are the set of \emph{primitive} variables used. 
The source function $\Sigma$ takes the form
\beq{ 
\Sigma \equiv \Theta + \frac{2 P X}{r} \, , \label{Sigma}
}
where
\beq{
\Theta  =  \alpha a \left[  \left( S v - E \right) \left( 
8 \pi r P  + \frac{m}{r^2} \right)
+ P  \frac{m}{r^2}  \right] \, . 
\label{Theta-both}
}

With this source, the governing equations with which we solve for our metric functions 
are the Hamiltonian constraint of the ADM \cite{adm} formulation
\beq{
\frac{ a^\prime }{a} = a^2 \left[ 4 \pi r  E  - \frac{1}{2 r} \right] + \frac{1}{2 r} \, ,
\label{polar-areal-hamiltonian-const}
}
and the polar-areal slicing condition
\beq{
\frac{\alpha '}{\alpha}
= a^2 \left[ 4 \pi r \left( S v + P \right) + \frac{m}{r^2} \right] \, .
\label{polar-areal-slicing-condition}
}
In order to monitor how near the spacetime is to forming an apparent horizon, we 
employ the mass aspect function,  $m$, 
\beq{  
m(r,t) \equiv \frac{r}{2} \left( 1 - 1 / a^2 \right) \, . \label{massaspect}
}
From Birkhoff's theorem, we know that an apparent horizon forms in the limit $2m/r\rightarrow~1$. 
We note that Eqs.~(\ref{polar-areal-hamiltonian-const}-\ref{polar-areal-slicing-condition}) 
were used
to compute the fluid equation source terms~(\ref{Sigma}-\ref{Theta-both}), and that flat space 
equations are obtained by 
setting $\Theta = 0$ and $X=1$. 

With the conservation equations (\ref{energy-cons})-(\ref{current-cons}), the equation of state (EOS) closes the 
system of hydrodynamic equations.  
Largely due to the extensive nature of our parameter space survey, 
we restrict the current study to continuum state equations (i.e.~we do not use tabulated
equations of state).
The polytropic EOS
\beq{
P = K \rho_\circ^\Gamma  \, , \label{polytrope-eos} 
}
is used only when calculating initial conditions for a star.  
Here, $K$ is taken to be constant (isentropic condition)
while $\Gamma$ is the adiabatic index.  After $t=0$, we allow for the development of shocks and therefore
only use the ``ideal-gas'' or ``gamma-law'' equation of state (\ref{ideal-eos}).
Our initial neutron star  models are approximated by solutions of the spherically-symmetric 
hydrostatic Einstein equations, called the Tolman-Oppenheimer-Volkoff (TOV) solutions 
\cite{oppenheimer-volkoff,tolman-paper,tolman-book}.  We simulate the stiffness of matter at 
super-nuclear densities
by setting $\Gamma=2$ in all of the calculations discussed here.
As pointed out by Cook et al. \cite{cook-shap-teuk-1992}, the constant $K$ can be 
thought of as the fundamental length scale of the system, which one can use to scale any 
dynamical quantity with set values of $(K,\Gamma)$ to a system with different values $(K',\Gamma')$.
As with $G$ and $c$, we set $K=1$, thereby making our equations dimensionless.  We note 
that for a specific value of $\Gamma$, the TOV solutions generically constitute a 
one-parameter family, where the central value of the fluid density, $\rho_c$, 
serves as a convenient parameter. 
The ADM masses of stars for a TOV family governed by the ideal-gas EOS typically depend 
on the value of $\rho_c$, with $M_{\rm ADM} \to 0$ for $\rho_c \to 0$, and $M_{\rm ADM}$ 
achieving a global maximum at some value $\rho_c = {\bar \rho}_c$.  Stars with 
$\rho_c < {\bar \rho}_c$ are stable against radial perturbations, while those 
with $\rho_c > {\bar \rho}_c$ are dynamically unstable in radial perturbation theory. 
In the experiments described below, we generally work with stars that have central 
densities significantly less than ${\bar \rho}_c$.

After the initial, star-like solution is calculated, an in-going velocity profile
is added to drive the star to collapse.  In order to do this, we follow
the prescription used in \cite{gourg2} and \cite{novak}.  The method entails specifying the 
coordinate velocity 
\beq{
  U \equiv \frac{ d r }{ d t } = \frac{ u^r }{ u^t }   \, ,
  \label{radial-coord-velocity}
}
of the star.  In general, the profile takes the algebraic form:
\beq{ 
U_g(x) = A_0 \left( x^3 - B_0 x \right) \, . \label{v-profile-general}
} 
The two profiles that were used in \cite{novak} are 
\beqa{
U_1(x) & = &   \frac{U_\mathrm{amp}}{2} \left( x^3 - 3 x \right) \, , \label{U1} \\ 
U_2(x) & = &   \frac{27 \, U_\mathrm{amp}}{10 \sqrt{5}} \left( x^3 - \frac{5 x}{3} \right)  
\, , \label{v-profile-12} 
} 
where $x \equiv r / R_\star$ and $R_\star$ is the radius of the TOV solution. Unless stated 
otherwise, the $U_1$ profile will be used for all the results herein. 
The velocity profile is added consistently to the TOV solution by 
recalculating $a$ and $\alpha$ via Eqs.~(\ref{polar-areal-hamiltonian-const}-\ref{polar-areal-slicing-condition}) once the profile has been assigned to a given star.
Further details concerning the calculation of initial data can be found
in~\cite{noble,noble-choptuik2}.

As mentioned above, 
previous critical phenomena studies of perfect fluids have focused on models governed by
the so-called ``ultrarelativistic'' EOS (\ref{ultra-eos}).
This can be thought of as an ultrarelativistic limit of~(\ref{ideal-eos}),
wherein $\rho_{\circ} \epsilon \gg \rho_{\circ}$ or $\rho \simeq \rho_{\circ} \epsilon$.
In this limit, $D$ becomes insignificant and one is left with two equations of motion for the fluid, which can 
be easily derived from Eqs.~(\ref{conservationeq})--(\ref{Theta-both}) by ignoring the EOM for $D$ and 
setting $E=\tau$.  The full expressions are given in~\cite{noble}.
In this paper, we only use the ultrarelativistic EOS in order to dynamically calculate 
ultrarelativistic Type~II critical solutions.   All other 
computations are performed using 
the ideal-gas EOS (\ref{ideal-eos}).

%%%%%%%%%%%%%%%%%%%%%%%%%%%%%%%%%%%%%%%%%%%%%%%%%%%%%%%%%%%%%%%%%%%%%%%%%%%%%%%%%%%%%
%%%%%%%%%%%%%%%%%%%%%%%%%%%%%%%%%%%%%%%%%%%%%%%%%%%%%%%%%%%%%%%%%%%%%%%%%%%%%%%%%%%%%
%%%%%%%%%%%%%%%%%%%%%%%%%%%%%%%%%%%%%%%%%%%%%%%%%%%%%%%%%%%%%%%%%%%%%%%%%%%%%%%%%%%%%
%% CHAPTER:   %%%%%%%%%%%%%%%%%%%%%%%%%%%%%%%%%%%%%%%%%%%%%%%%%%%%%%%%%%%%%%%
%%%%%%%%%%%%%%%%%%%%%%%%%%%%%%%%%%%%%%%%%%%%%%%%%%%%%%%%%%%%%%%%%%%%%%%%%%%%%%%%%%%%%
\section{Numerical Techniques \& Computational Issues}
\label{sec:numerical-techniques}

In this section we discuss the numerical techniques we use
to simulate the highly-relativistic flows encountered in the driven
collapse of neutron stars.  The simulations entail solution of a system of coupled, partial and ordinary 
differential equations that describe how the fluid and gravitational field evolve in time.  
In section~\ref{sec:instability} we also discuss an instability that generically 
appears in our calculations that are very close to the black hole threshold---this 
instability ultimately limits how closely we can tune any given family parameter to 
criticality.

We use the Rapid Numerical Prototyping Language (RNPL)~\cite{marsa-choptuik}
to handle check-pointing, input/output, and memory management for all our simulations. Secondary 
routines are called to solve the fluid and geometric equations.  We use second-order high-resolution 
shock-capturing (HRSC) methods to evolve the fluid.  The discrete equations 
are derived using a 
finite volume approach and are detailed in Appendix~\ref{app:fluid-methods}.  
We generally use a Roe-type method \cite{roe-1981,romero} as our approximate 
Riemann solver.  However, particularly in the investigation of the instability 
mentioned above, we have sometimes used 
the Marquina flux formula \cite{donat-marquina} and  
Harten and Hyman's \cite{harten-hyman} entropy-fix for Roe's method,
to compare with the basic Roe solver.
For accurate and stable resolution of shocks, 
we use the \textrm{minmod} 
slope-limiter \cite{vanleer-1979} to reconstruct the primitive variables at cell interfaces.  
We have also implemented the the linear \textrm{MC} \cite{vanleer-1977} and \textrm{Superbee} \cite{roe-1985} 
limiters, as well as the high-order essentially non-oscillatory (ENO) scheme \cite{shu1997}, 
but have found the \textrm{minmod} limiter  to provide 
the most stable evolutions near the threshold of black hole formation while still resolving shocks adequately.

In order to track the continuously decreasing spatio-temporal scales typically seen in CSS phenomena, 
we use a nonuniform grid that refines as needed.  The origin, the point upon which the matter collapses,
is the natural setting for the smallest dynamical scales and we therefore only refine the innermost 
region.  Our implementation was inspired by Neilsen \cite{neilsen-thesis} and is detailed in \cite{noble}. 
The basic idea is to segment the discrete domain into 
three regions: an innermost uniform grid with the smallest 
grid spacing, $\Delta r_a$, composed of $N_a$ cells,  an adjacent intermediate grid with $N_b$ 
cells of sizes $\Delta r \propto r$,  and an outermost uniform grid of $N_c$ cells.  Refinement 
occurs when the maximum of $2m(r,t)/r$ is attained within
$N_a/2$ cell widths from the origin.  Interpolation is 
performed with a $3^\mathrm{rd}$-order ENO interpolation procedure written by Olabarrieta \cite{olabarrieta}. 
We usually set $N_a \simeq 300-600$, $N_b \simeq 2 N_a$, $N_c \simeq 10-20$ and 
adopt an initial value for $\Delta r_a$ 
such that the outer boundary lies at about $5$--$10$ times the initial radius of the star. 

Time integration is performed separately from the spatial discretization using the 
method of lines. 
Specifically, 
an explicit, two-step predictor-corrector technique, called Huen's method,
is used to time-advance the 
ODEs that result from the spatial discretization of our time dependent PDEs.
Discrete timesteps, $\Delta t$, for the ODE integration are constrained to 
magnitudes given by $\Delta t / \Delta r_a < 0.4$, ensuring that the Courant-Friedrichs-Levy condition
for our scheme is not violated.
Additional details concerning the time integration are given in
App.~\ref{app:fluid-methods} as well as in~\cite{noble}.
Results from a shock tube test, which 
measures our code's ability to evolve discontinuities, and a convergence test involving 
the evolution of a self-gravitating distribution of fluid are presented in 
App.~\ref{app:numerical-tests}.

%%%%%%%%%%%%%%%%%%%%%%%%%%%%%%%%%%%%%%%%%%%%%%%%%%%%%%%%%%%%%%%%%%%%
%%%%%%%%%%%%%%%%%%%%%%%%%%%%%%%%%%%%%%%%%%%%%%%%%%%%%%%%%%%%%%%%%%%%
\subsection{Primitive Variable Calculation}
\label{sec:prim-var-calc}

Since only the conserved variables are evolved by the HRSC schemes discussed above, the primitive 
variables must be derived
from the conserved variables after each predictor and corrector step in order to 
compute fluxes $\mathbf{f}$ and source functions $\bm{\psi}$ for the next evolution step.  
This involves inverting the three equations 
$\mathbf{q}=\mathbf{q}(\mathbf{w})$---given by the definitions of the conserved
variables (\ref{D-S-tau-E})---for the three unknown primitive variables,
$\mathbf{w}$.
While closed-form expressions for the inverted equations exist for the ideal-gas EOS, 
numerically solving the equations is far more efficient \cite{eulderink}.
At each grid point, we use a Newton-Raphson method to find the values of $\mathbf{w}$ 
that minimize the residuals of the conserved variable 
definitions (\ref{D-S-tau-E}).  Instead of solving the full $3$-by-$3$ 
system at each point, an identity function $\mathcal{I}$---derived 
from Eq.~(\ref{D-S-tau-E})---is
used as a residual, making the solution process one-dimensional.  This makes the procedure
much more efficient.

Our method for performing the inversion,
which is discussed further in App.~\ref{app:pseudo-code-prim},
is based on one specially suited for 
spherically symmetric ultrarelativistic flows~\cite{neilsen-thesis,neilsen}, and
uses a residual function based on the definition of $E$:
\beq{
\mathcal{I}(H) \ = \ H W^2 - \tau - D - P   \, . \label{pvar-resid2}
}
Here, $H(r,t) = \rho_\circ h$ is the enthalpy.  
In order to increase the accuracy of our computation of $\mathbf{w}$, we use
different 
methods for calculating the residual $\mathcal{I}$ and its 
derivative $\mathcal{I}^\prime = \partial \mathcal{I} / \partial H$ in different 
regimes (including both the ultrarelativistic and nonrelativistic limits).
The ``nonrelativistic'' and 
``intermediate'' methods originated from~\cite{neilsen-thesis,neilsen}, where flows in the 
ultrarelativistic limit were also studied.  However, we have found that in the 
ultrarelativistic limit, where $\Lambda = S/H \rightarrow \infty$, the intermediate method 
still gives imprecise results.  This imprecision can be traced 
to a loss of precision in the calculation of the quantity
\beq{
1 - v\left(\Lambda\right) = 1 - \sqrt{1+1/4\Lambda^2} + 1/2\left|\Lambda\right|  \, . 
\label{inter-imprecision}
}
We thus expand all nonlinear expressions appearing in the conversion to primitive 
variables in powers of 
$b=1/2\left|\Lambda\right|$, which yields results with increased floating-point accuracy.
In the other limit, $|\Lambda| \ll 1$, where the 
the system is 
nonrelativistic, we use expansions up 
to $O(\Lambda^9)$ that similarly reduce the influence of round-off errors.
In practice, the ultrarelativistic regime 
is defined by an adjustable parameter $\Lambda_\mathrm{High}$ and the nonrelativistic regime by
$\Lambda_\mathrm{Low}$.  For example, for all the results shown below, we used 
$\Lambda_\mathrm{High} = 10^2$  and $\Lambda_\mathrm{Low} = 10^{-4}$; these values ensure 
that the leading-order error terms in the ultrarelativistic and nonrelativistic 
expansions are below the intrinsic round-off error of the computations.

\begin{figure}[htb]
\includegraphics[scale=0.42]{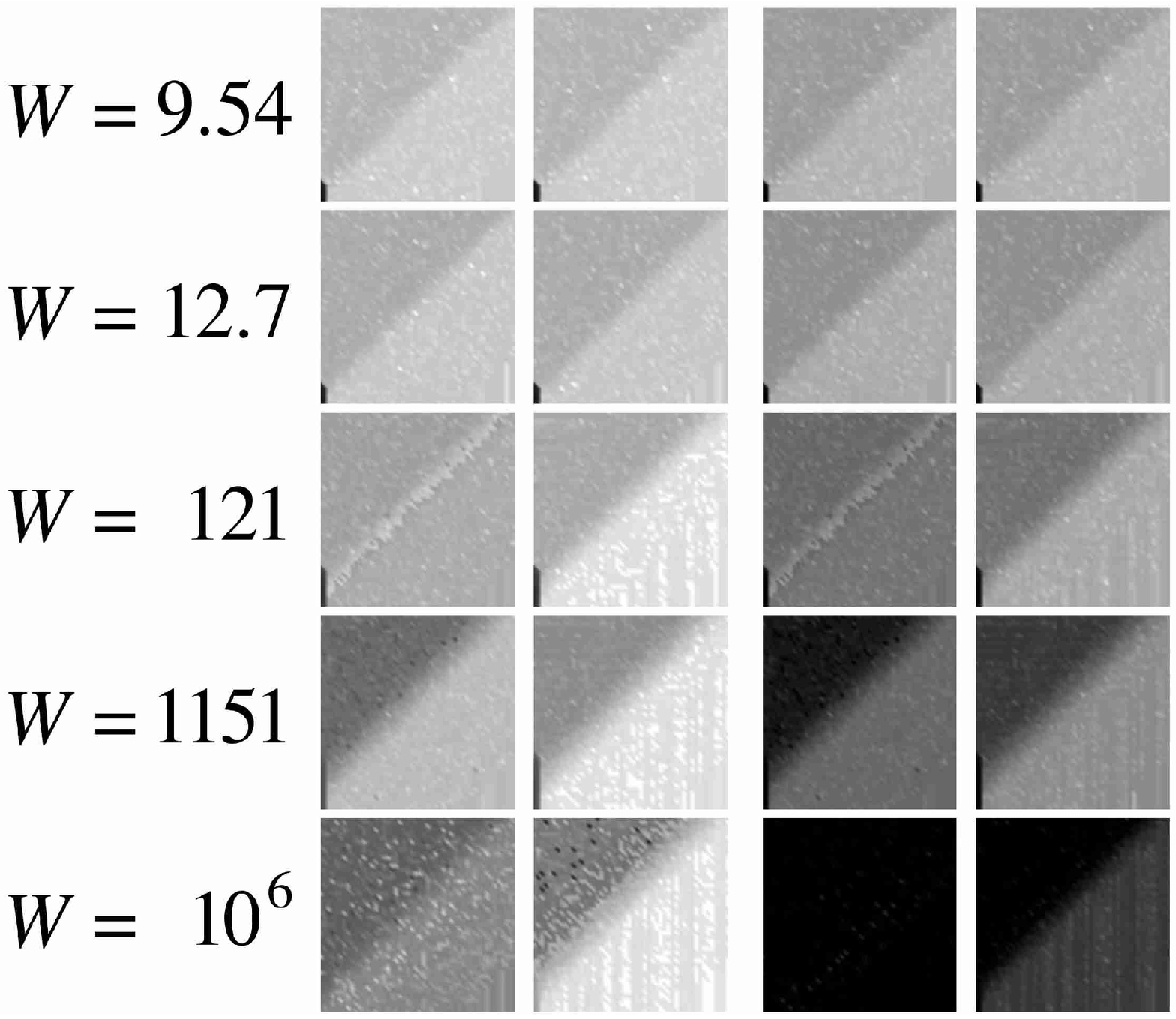}
\includegraphics[scale=0.45]{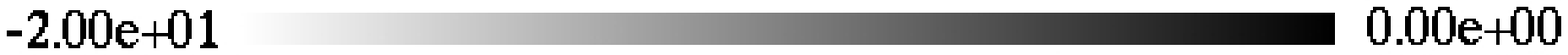}
\caption{Comparison of the accuracy achieved with our primitive 
variable solver and that used in \cite{neilsen}.  The first (third) column shows 
$\log_{10}$ of the relative error between the exact value of $v$ ($\rho_\circ$) and the value  obtained 
using the method described in~\cite{neilsen}, while the second (fourth) column shows $\log_{10}$ of the relative error 
between the exact value of $v$ ($\rho_\circ$) and the value obtained from the method described in 
Sec.~\ref{sec:prim-var-calc}.  In any given plot, the vertical (horizontal) axis is discretized into 
50 uniformly-spaced values of $\log_{10}(P)$ ($\log_{10}(\rho_\circ)$); the minimum (maximum) value used for  
both $P$ and $\rho_\circ$ is $2.5 \times 10^{-17}$ ($10^6$) .  Each row contains plots of $(P,\rho_\circ)$-space
calculations that were performed with a given value of $W$ shown to the left of the row.  
The uniformly-spaced color map is shown at the bottom and indicates
that the maximum (darkest shading) represents errors comparable to and exceeding $100\%$, 
while the minimum (lightest shading) represents errors comparable to and below machine precision.  
\label{fig:prim-solver-survey}}
\end{figure}

Comparisons of the accuracy of our improved method and the work presented
in~\cite{neilsen-thesis,neilsen} are shown in 
Fig.~\ref{fig:prim-solver-survey}.  In order to estimate the relative error in the
primitive  
variable calculation for each method, we seed each solver with guesses for 
$\mathbf{w}$ that are a fixed factor away from 
the true solution, $w^{(k)}_\mathrm{guess} = w^{(k)}_\mathrm{exact} \left( 1 + z^{(k)} \right)$, 
where $\mathbf{w}_\mathrm{exact}$ is the exact solution, and $w^{(k)}$ denotes the $k$-th
component of the state vector $\mathbf{w}$.
A constant set of seeds $z^{(k)} = \left\{ - 0.2121, - 0.0208941, - 0.25971 \right\}$ are 
used in order 
to put all calculations on equal footing.  
Although the convergence of both methods {\em does} depend on the size of the 
$z^{(k)}$, our pre-specified initial values are generally further from the true solution
values than they are in the context of an actual calculation.
In addition, 
use of different seeds with magnitudes comparable to those given above yielded similar 
results, suggesting that these particular values of $z^{(k)}$ are
appropriately representative. 
For each method, once a best estimate for $\mathbf{w}$ is computed, the 
relative error for the solver is computed as  
$\left( \mathbf{w} - \mathbf{w}_\mathrm{exact}\right)/\mathbf{w}_\mathrm{exact}$. 
Fig.~\ref{fig:prim-solver-survey} shows the logarithm of the relative errors for 
the variables $v$ and $\rho_\circ$ (the trend of the error in $P$ is similar to 
that of $\rho_\circ$). 
The improved accuracy of our new method is most noticeable in the 
computation of $v$, as demonstrated by 
the first two columns in the figure.  We note that our method can accurately 
calculate $\mathbf{w}$ 
for $W > 10^3$ and for all $P,\rho_\circ$ tested, while the method 
described in~\cite{neilsen} develops 
significant problems when $W \gtrsim 10^3$ and $P > \rho_\circ$.  

Even though the above methods improved the accuracy of the primitive variable calculation, 
significant errors still remain for highly-relativistic flows ($W > 10^5$) where $P$ 
and $\rho_\circ$ differ by orders of magnitude---i.e.~when 
$P \gg \rho_\circ$  or $\rho_\circ \gg P$.  In these regimes, machine precision limits the 
accuracy of the calculation since terms in $\mathcal{I}$ and $\mathcal{I}^\prime$ 
become numerically insignificant (relative to other terms)
even though their presence is essential to the computation of a solution.
The effect from round-off error in these regimes can easily be seen by plotting $\mathcal{I}(H)$
using different orders of numerical precision, which we have done using arbitrary precision arithmetic 
in \texttt{Maple}. In order to accurately calculate $\mathbf{w}$ in these regimes, 
one would need a new algorithm that performs better
in these regimes, or use higher-precision arithmetic in the simulation.  Fortunately, the improvements 
we have made seem to be sufficient for our current purposes.

%%%%%%%%%%%%%%%%%%%%%%%%%%%%%%%%%%%%%%%%%%%%%%%%%%%%%%%%%%%%%%%%%%%%
%%%%%%%%%%%%%%%%%%%%%%%%%%%%%%%%%%%%%%%%%%%%%%%%%%%%%%%%%%%%%%%%%%%%
\subsection{Instability at the Sonic Point in the CSS Regime}
\label{sec:instability}

In this section we provide a description of an instability 
that develops in our calculations in the vicinity of 
the sonic point in near-critical evolutions.  
When using the unmodified approximate Roe solver,
this instability made it impossible for us to
obtain consistent brackets about the critical parameter, $p^\star$, 
for $|p-p^\star| \lesssim 10^{-9}$.
This significantly hindered
our study, since we found that we needed to tune quite closely to the threshold solution
in order to calculate an accurate value of the scaling exponent $\gamma$.

\begin{figure}[h]
\includegraphics[scale=0.4]{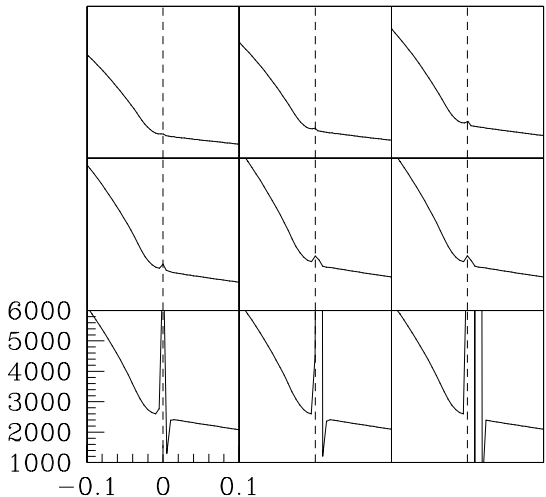}
\caption{Conserved variable $D(\mathcal{X},\mathcal{T})$ from the most nearly critical evolution 
obtained with the use of the approximate Roe solver without smoothing. 
$\mathcal{X}$ and $\mathcal{T}$ are defined by Eqs.~(\ref{X-ss}) and (\ref{T-ss}), 
respectively.  
The dashed line indicates the location of the sonic point, $\mathcal{X}=0$.  No refinement takes place 
during the period shown here,  and $\Delta r_a \simeq1.55\times10^{-7}$.
From left-to-right and top-to-bottom, the $\mathcal{T}$ values of the frames are
$-10.4109$, $-10.4977$, $-10.5916$, $-10.6938$, $-10.7822$, $-10.7823$, 
$-10.7824$, $-10.7825$, $-10.7826$.
The evolution started with a TOV star of central density $\rho_c = 0.05$ 
that was perturbed using profile $U_1$ (see~(\ref{U1})) 
with the overall amplitude factor, $U_{\rm amp}$, tuned to produce near-critical
evolution.
\label{fig:roe-q1-instability}}
\end{figure}

An example of the instability seen in evolutions using primitive 
variable reconstruction is shown in Fig.~\ref{fig:roe-q1-instability}.  The conserved 
variable $D$ is plotted in CSS coordinates ${\cal T}$ and {\cal X} defined later in 
the paper (Eqs.~(\ref{T-ss}) and (\ref{X-ss}), respectively). 
The last five frames show data from the last 5 time steps before the code crashes, 
while the 
first four frames are more distributed in $\mathcal{T}$.  
Hence, we see that the feature at the 
sonic point exists for many discrete time steps before its growth produces a 
code crash.

The instability manifests itself in different ways, depending on the type of 
cell reconstruction used.  For example when using the conserved variables to 
reconstruct the solution at the cell borders, we find that the conserved variables themselves
remain smooth, but that each of the primitive variables exhibits
persistent oscillations near the sonic point that span of order 2-4 grid cells.
On the other hand,
reconstructing with the primitive variables leads to smooth $\mathbf{w}$ but oscillations in $\mathbf{q}$. 
A third and final reconstruction method was tried with the so-called characteristic
variables, which are the advected quantities in the equations found by
diagonalizing the quasi-linear form 
of Eq.~(\ref{conservationeq}).  This method was significantly more diffusive but 
{\em less} stable
than reconstruction with primitive variables. 

The instability is also sensitive to the slope limiter used, 
with Superbee and Monotonized Central-differenced (MC) 
limiters producing more spurious oscillations than the more diffusive minmod limiter. 
We also found no improvements by varying the order of the ENO reconstruction from
$O(\Delta r^2)$ through $O(\Delta r^{10})$.

In terms of ruling out potential sources of the problem, 
we {\em have} ensured that the regridding procedure is not responsible for 
the instability. 
To accomplish this, we first evolved a system that was tuned near the critical solution.  We extracted 
the grid functions at a specific time, ${\bar t}$, before the 
appearance of instability, and interpolated them  onto a new grid fine enough so that no 
further refinement would be required in the subsequent evolution.
The data was allowed to evolve from this 
time, and the instability developed in the same manner and at the same time, ${\bar t}$,
as in the original run.

Moreover, we find that the instability does not ``converge away.''  We tuned the initial data 
towards criticality for three different levels of refinement, where refinement was done locally so that 
$\Delta r_{l}(r) = 2 \Delta r_{l+1}(r)$ for all $r$, and $l$ is the ``level'' of refinement.  
We find that as 
$l$ increases the oscillations associated with the instability do not significantly change 
in magnitude and 
remain confined to approximately the same number of grid cells.  Also, the solutions eventually 
diverge at the sonic point in all cases. 

In order to describe the likely source of the instability, we first need to provide a better 
description of the near-critical solution.  
When the initial data has been tuned close to the critical solution at the threshold of black hole
formation, the behavior near the origin is self-similar up to 
the sonic point, $r_s$, where the flow velocity equals the speed of sound, $c_s$ (\ref{ideal-eos-quantities}).  
For these solutions the fluid becomes 
ultrarelativistic---e.g. $P \gg \rho_\circ$---for $r<r_s$ and we expect
that $c_s(r) \rightarrow 1 $ in that region.  Also, from previous 
ultrarelativistic studies using
$\Gamma=2$ such as \cite{brady_etal,neilsen-crit}, we 
expect that $v \rightarrow 1$  for $r > r_s$.  Thus, about the sonic point, 
the characteristic speeds (\ref{ideal-piphi-evalues}) should take the values given 
in Table~\ref{table:ultrarel-char-speeds}.  

\begin{table}
\caption{Asymptotic values of the fluid's characteristic speeds in the ultrarelativistic 
limit. The sonic point is located at $r = r_s$. }
\label{table:ultrarel-char-speeds}  
\begin{ruledtabular}
\begin{tabular}{ccc}
Characteristic Speed  &$\lambda(r <  r_s)$ &$\lambda(r > r_s)$  \\
\hline
$\lambda_1$  &$ < 1$   & $\sim 1$ \\
$\lambda_+$  &$\sim 1$   & $\sim 1$ \\
$\lambda_-$  &$\sim -1$   & $\sim 1$ \\
\end{tabular}
\end{ruledtabular}
\end{table}

\begin{figure}[htb]
\includegraphics[scale=0.4]{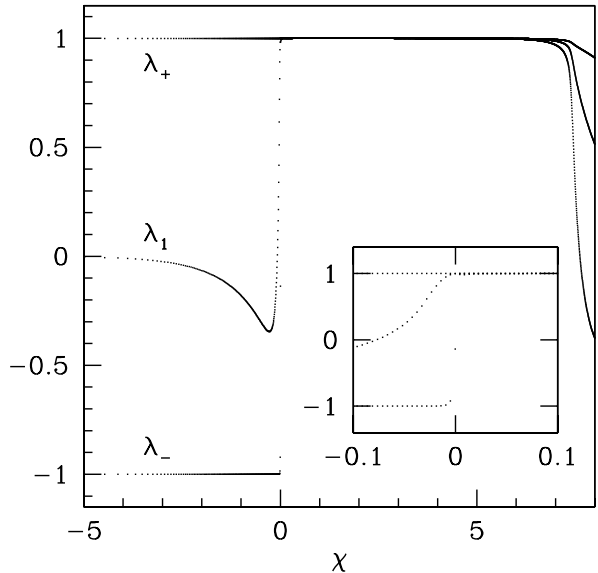}
\caption{Characteristic speeds of the fluid for the most 
nearly critical solution obtained with the approximate Roe solver without smoothing.  The wave speeds 
are plotted here as functions of the self-similar coordinate $\mathcal{X}$, and are shown at 
$\mathcal{T}=-10.6938$.
A closer view of the characteristic speeds near the sonic point is shown as an inset in the lower-right of 
the plot, revealing the severity of the discontinuity in $\lambda_-$ as discussed in the text.
\label{fig:crit-characteristics}}
\end{figure}

\begin{figure}[htb]
\includegraphics[scale=0.4]{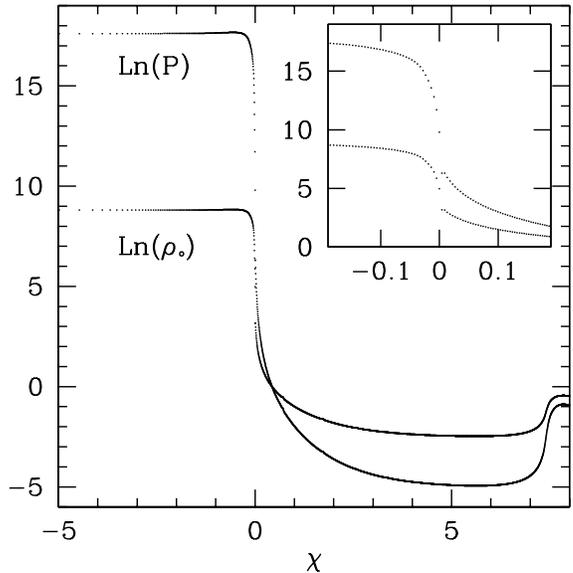}
\caption{Pressure and rest-mass density of the most nearly critical solution obtained with
the approximate Roe solver without smoothing.  
Both quantities are plotted as a function of the 
self-similar coordinate $\mathcal{X}$, and are shown at $\mathcal{T}=-10.6938$.
Interior to the sonic point, $\mathcal{X}=0$, the fluid is clearly in 
the ultrarelativistic limit,
with $P / \rho_\circ \approx 10^4$. 
However, beyond the sonic point, the flow is {\em not} ultrarelativistic---in fact 
$P < \rho_\circ$ in most of the domain exterior to $\mathcal{X}=0$. 
A closer view of the distributions near the sonic point is shown in the inset,
and more clearly illustrates the formation of an expansion shock as discussed in the text.
\label{fig:p-rho-ultra-regime}}
\end{figure}

In fact, this is exactly what we find when using the ideal-gas state equation, as seen 
in Fig.~\ref{fig:crit-characteristics}
and Fig.~\ref{fig:p-rho-ultra-regime}.  In Fig.~\ref{fig:p-rho-ultra-regime}  we
see that $P \gg \rho_\circ$ within the self-similar region, but that $P(r) < \rho_\circ(r)$  for 
$r > r_s$.  

From these plots we also find that the transition 
from the ultrarelativistic regime to the exterior solution---defined by $r>r_s$, and 
characterized by an absence of self-similarity and decay to asymptotic flatness---is
quite abrupt.
For instance, the discontinuity in $\lambda_-$ is resolved by only a few grid points, 
signifying the presence of a shock which can also be 
seen for $r \sim r_s$ in the plots of $P(r)$ and $\rho_\circ(r)$ shown in Fig.~\ref{fig:p-rho-ultra-regime}.  
Since $\lambda_-(r<r_s)<0$ and $\lambda_-(r>r_s)>0$,
the discontinuity  represents a point of transonic rarefaction.  Also, the shock appears to 
be an expansion shock, which is entropy-violating, since it travels into a region of higher pressure and density. 
The reason why we can have an entropy-violating 
shock  develop is because it is coincident with a change in curvature:  as the high-pressure 
matter leaves the confines of the potential, it freely expands and its internal energy is converted 
into bulk kinetic energy.  

LeVeque states in \cite{leveque1} that the Roe solver can lead to the wrong Riemann solution 
at transonic rarefactions (in flat spacetime) since the linearization that the Roe solver performs 
on the EOM leads to a Riemann solution with only discontinuities and no rarefaction waves.
He illustrates this point in \cite{leveque2} using a boosted shock tube test that makes the rarefaction transonic.
Other failures of Roe's method that are attributed to its linearization have been 
shown by Quirk \cite{quirk}, and by Donat et al. 
\cite{donat-font} where an unphysical ``carbuncle'' forms in front of a relativistic,
supersonic jet.  

\begin{figure}[htb]
\includegraphics[scale=0.4]{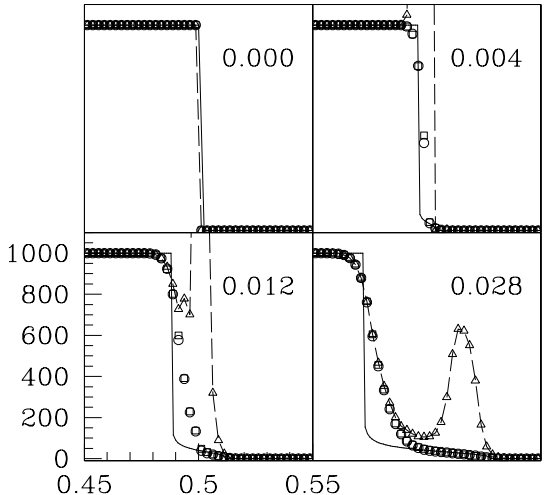}
\caption{One-dimensional, 
slab-symmetric shock tube test to simulate the discontinuity observed 
near the sonic point of near-critical solutions.  The rest-mass 
density $\rho_\circ(x,t)$ computed using different Riemann solvers is plotted versus $x$ in each constant-$t$ frame.  
Solution time is shown in the upper-right part
of each frame.  The solid line without points is an exact Riemann solution, the dashed line with triangles
corresponds to the solution obtained with the approximate Roe solver, squares represent the solution from
Marquina's method, and circles are from the Roe scheme with entropy fix. 
\label{fig:css-shocktube}}
\end{figure}

To see whether Roe's method contributes to the instability, we implemented the Marquina method and 
the entropy-fix for the Roe scheme due to Harten and Hyman~\cite{harten-hyman}.  
A comparison between the three methods is shown in Fig.~\ref{fig:css-shocktube}, 
where we have evolved a shock tube problem that emulates the fluid state about the sonic point
of near-critical solutions.  The initial conditions used for this test are 
$\left\{\rho_L, v_L, P_L\right\}=\left\{1.0\times 10^3 , -0.3, 1.0\times 10^6\right\}$ and
$\left\{\rho_R, v_R, P_R\right\}=\left\{0.3, 0.9994, 1.0\right\}$. These values are such 
that, initially, ${\lambda_+}_L\simeq0.9995$, ${\lambda_+}_R\simeq0.99998$, 
${\lambda_-}_L\simeq-0.99987$, and ${\lambda_-}_R\simeq0.98296$, closely 
approximating the values listed
in Table~\ref{table:ultrarel-char-speeds}.  
The Roe, Marquina and Roe with entropy-fix evolutions each used 400 points in the entire grid 
of domain $0 \le x \le 1$ (only a portion of the grid is shown here), and both used a 
Courant factor of $0.4$.  The exact solution was
obtained from the Riemann solver provided in \cite{marti-muller-living} with $1000$ points, 
using the same range in $x$ and same initial conditions as the Roe and Marquina runs.  
The Marquina method and entropy-fixed Roe method produce similar---yet more diffused---solutions than 
the exact solver, while  the basic Roe solver severely diverges from the exact solution near the 
transonic rarefaction during the first few time steps.  Even though the initial divergence of the Roe solution 
eventually vanishes, a relic feature still exists and propagates away from the center.
If we were to reverse the evolution of the Roe solution shown here, the sequence would be reminiscent 
of how the instability in $D$ grows near the sonic point
of near-critical solutions (Fig.~\ref{fig:roe-q1-instability}).

This shock tube test suggests that our use of Roe's method may underlie the 
observed instability.  
In order to address this possibility, we implemented both the 
Marquina solver and the entropy-fix in the general relativistic code and 
tuned towards the critical solution.  We were able to tune to 
$|p-p^\star| \approx 5 \times 10^{-11}$ with Marquina's method, 
which is approximately a factor $10^2$ closer to $p^\star$ than
we reached with Roe's method.  Also, the ``bump'' at $r_s$ developed at a later $\mathcal{T}$ 
with Marquina's method.  However, the use of 
Marquina's flux formula did not completely solve the problem since evolutions using it also 
eventually succumb to the instability, preventing us from tuning beyond 
$|p-p^\star| \approx 5 \times 10^{-11}$.  Moreover, Roe's method with the entropy-fix, 
performed more poorly than when used without the entropy-fix, 
impeding any further tuning  past $|p-p^\star| \approx 2 \times 10^{-8}$.

A possible explanation for the failure of the entropy-fixing methods to stabilize 
near-critical evolutions may involve  the fact that they {\em do} provide an entropy-fix.  
As mentioned before, these methods would evolve the critical solution's 
expansion shock to a rarefaction fan if no source was present (see Fig.~\ref{fig:css-shocktube}). 
However, the source in the EOM and/or the geometric factor in the flux must be what 
keeps the expansion shock from being dissipated away since the shock is a part of the critical solution. 
The competition between the approximate Riemann solutions of the entropy-fixing methods and the 
spacetime may exacerbate the instability. 
However, it is unclear why Marquina's method is more successful than Roe's method with 
an entropy-fix.

Even though Marquina's method provides a marked improvement over Roe's method, 
it does not eliminate the sonic-point instability.
A first attempt to explicitly dissipate the instability involves applying artificial viscosity 
in the region.  We follow Wilson's \cite{wilson2} artificial viscosity method and set 
$P \rightarrow P + Q$ in $\mathbf{f}$ alone, where 
\beq{
Q \ = \ c_\mathrm{av} D \, \left(\Delta r \pderiv{v}{r} \right)^2 \, . \label{av-term}
}
and $c_\mathrm{av}$ is a user-specified parameter.  
However, the instability worsens as $v \rightarrow 1$, 
and, since $Q$ is not proportional to $W$ like other terms in $\mathbf{f}$,
$Q$ becomes irrelevant as the flow becomes
more relativistic. 
As a result, moderate values of $c_\mathrm{av}$ lead to insignificant changes in behavior near the sonic point, while 
extremely large values tend to \emph{amplify} the instability and induce additional 
high-frequency modes near $r_s$. 

Since we find that the instability near $r_s$ becomes more severe as 
$\lambda_-$ became more discontinuous, 
our second attempt to control the blow-up
entails smoothing the conserved variables about $r_s$ at every 
predictor/corrector step of the fluid update.  The smoothing is done once the $\lambda_-$ discontinuity
develops.  Specifically, we start to use the smoothing procedure when $|p-p^\star| \le 10^{-8} - 10^{-9}$, and at
times when $\lambda_-$ begins to be resolved over approximately $10$ or fewer zones.  We can use
the same time, $t_s$, to begin smoothing for all runs since the evolution for $t<t_s$ is almost identical 
for all near-critical values of $p$.  The smoothing is performed over the first contiguous set of 
points, $r^{\mathrm{sm}}_j$, that satisfy 
$-\lambda_-^{\min} \ < \ \lambda_-(r^{\mathrm{sm}}_j) \ < \ \lambda_-^{\min}\ $, where $\lambda_-^{\min}$ is some 
 adjustable parameter which we set to $0.95$.   
The smoothing operation replaces the quantity $q_j$ with $(q_{j+1} + q_{j-1})/2$.

We also find that the 
instability worsens as the number of points between the origin and the sonic point decreases, 
as occurs in those cases where the solution disperses from the origin instead of forming a black hole.  
Our ability to follow evolutions through to dispersal is necessary for calculation of 
the scaling exponent, $\gamma$, since, as described in the next section, one way 
of estimating $\gamma$ involves measuring the scaling of the 
global maximum, $T_{\rm max}$, of the stress energy trace, $T(r,t) \equiv {T^a}_{a}$, as a function
of $|p - p^\star|$.  We find that $T_{\rm max}$ is generally attained at a time when 
the fluid is beginning to disperse. 
Consequently, we found it necessary to refine the grid whenever the 
discontinuity or $\max_r\left(2m/r\right)$ reaches $r\sim r_a/2$. 

The diffusion introduced by the smoothing allows us to further tune toward the critical
solution, eventually to $|p-p^\star| \simeq 5 \times 10^{-12}$.  However, we are still 
unable to calculate the global maximum of $T$, $T_{\max}$, for the most nearly critical runs even though we can
identify them as being dispersal cases.  The minimum value of $|p-p^\star|$  for which 
we can calculate $T_{\max}$ is about $5 \times 10^{-10}$, as illustrated
in Fig.~\ref{fig:sub-scaling-alldata}.  This is far smaller, however, than 
we can achieve without smoothing or with any other method we have tried.
Surprisingly, smoothing $\mathbf{q}$ about the sonic point did not make the Marquina evolutions any more stable.

%%%%%%%%%%%%%%%%%%%%%%%%%%%%%%%%%%%%%%%%%%%%%%%%%%%%%%%%%%%%%%%%%%%%
%%%%%%%%%%%%%%%%%%%%%%%%%%%%%%%%%%%%%%%%%%%%%%%%%%%%%%%%%%%%%%%%%%%%
%%%%%%%%%%%%%%%%%%%%%%%%%%%%%%%%%%%%%%%%%%%%%%%%%%%%%%%%%%%%%%%%%%%%
\section{Simulation Results}
\label{sec:type-ii-critical}

In this section, we study the Type~II, CSS critical solution found at the black hole-forming threshold
of the parameter space described in \cite{noble-choptuik2}.
If not otherwise stated, the results in this section use $U_1$ (\ref{v-profile-12}) 
for the initial velocity profile, and the overall amplitude, $U_\mathrm{amp}$ of the 
velocity profile for the tuning parameter, $p$.
As previously mentioned, the stars that we able to drive to a Type~II black hole threshold 
generally have central densities, $\rho_c$, significantly smaller than the maximum 
value, ${\bar \rho}_c$, along the stable branch, which for $\Gamma = 2$ is $0.32$ in our 
units.
Although we were generally able to form black holes from stars with 
an initial rest-mass central density 
greater than $\rho_c^{\min} = 0.007$, we have closely tuned towards critical solutions
for only a handful of such initial states.  
(We were unable to form black holes from 
stars with $\rho_c < \rho_c^{\min}$ using an 
initially-ingoing velocity profile.)
In Table~\ref{table:type-ii-stars}, we list the 
central densities of the stars 
for which Type~II behavior was actually observed, and quantify how close to the critical value 
we were able to tune.
The instability described in Sec.~\ref{sec:instability} limited the tuning in all instances.  

\begin{table}
\caption{Star solutions in which we observed Type~II behavior, and the 
minimum black hole masses we were able to form from them. 
We denote the mass of the smallest black hole found for a given $\rho_c$ by 
${\min}(M_\mathrm{BH})$, $M_\star=M_\star(\rho_c)$ is the mass of the initial star solution, 
and ${\min}\left|p-p^\star\right|/p$ is the relative precision reached in $p^\star$ per star. 
The final columns lists the critical parameter values we obtained. }
\label{table:type-ii-stars}
\begin{ruledtabular}
\begin{tabular}{cccc}
$\rho_c$   &${\min}(M_\mathrm{BH})/M_\star$ &${\min}\left|p-p^\star\right|/p$ &$p^\star$\\
\hline
$0.01$  &$1\times 10^{-6}$  &$2\times 10^{-9}$   &0.889\\
$0.02$  &$6\times 10^{-7}$  &$1\times 10^{-9}$   &0.746\\
$0.03$  &$3\times 10^{-7}$  &$5\times 10^{-10}$  &0.634\\
$0.04$  &$6\times 10^{-8}$  &$2\times 10^{-11}$  &0.543\\
$0.05$  &$2\times 10^{-8}$  &$6\times 10^{-12}$  &0.469\\
%
%$0.01$  &$1\times 10^{-6}$  &$2\times 10^{-9}$   &0.88942207\\
%$0.02$  &$6\times 10^{-7}$  &$1\times 10^{-9}$   &0.74611650\\
%$0.03$  &$3\times 10^{-7}$  &$5\times 10^{-10}$  &0.633712118\\
%$0.04$  &$6\times 10^{-8}$  &$2\times 10^{-11}$  &0.543143513\\
%$0.05$  &$2\times 10^{-8}$  &$6\times 10^{-12}$  &0.46875367383\\
\end{tabular}
\end{ruledtabular}
\end{table}

From Table~\ref{table:type-ii-stars} it is clear that the instability's effect on 
our ability to find the critical parameter increases with decreasing $\rho_c$.  This is most likely 
due to the fact that sparser stars require greater in-going velocities in order to collapse, giving rise to 
more relativistic and, consequently, less stable evolutions.  We note, however, that our results represent 
great improvement over the precision obtained in \cite{novak}; the smallest 
black hole attained in that study was ${\min}(M_\mathrm{BH})/M_\star \sim 10^{-2}$. 
The success of our code is most likely due to our use of adaptive/variable mesh procedures and the great 
lengths we went to combat the sonic point instability.  

Unless otherwise stated, and for the remainder of the section, we focus on behavior
seen with the star having central density $\rho_c=0.05$.

\begin{figure}[htb]
\includegraphics[scale=0.4]{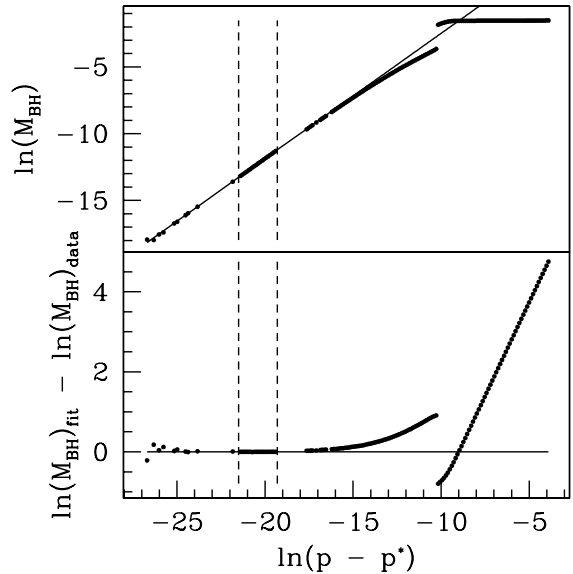}
\caption{Scaling behavior for supercritical---or black hole forming---solutions.
The top plot illustrates how the points from a series of supercritical runs follow the scaling 
law for the black hole mass (\ref{mass-scaling}), while the bottom plot 
shows how the data deviate from our best fit to this scaling law.  The two dotted 
lines delineate the data used in making the best fit; this data is plotted separately in
Fig.~\ref{fig:super-scaling-bestfit}.  Black holes were assumed to have formed when 
${\max}_r \, (2m(r,t)/r) \ge 0.995$.  The gaps between some of the points represent those 
runs that crashed before ${\max}_r\,(2m/r)$ reached this value.  Smoothing was used for 
$\ln{|p-p^\star|} < -19.3$, which is also where we start our fit. 
These runs used $\rho_c = 0.05$, $U = U_1$ and an initial grid defined by
$\{N_a, N_b, N_c, \Delta r_a, \mathrm{level}\} = \{300, 500, 20, 0.005, 0\}$.  
\label{fig:super-scaling-alldata}}
\end{figure}

\begin{figure}[htb]
\includegraphics[scale=0.4]{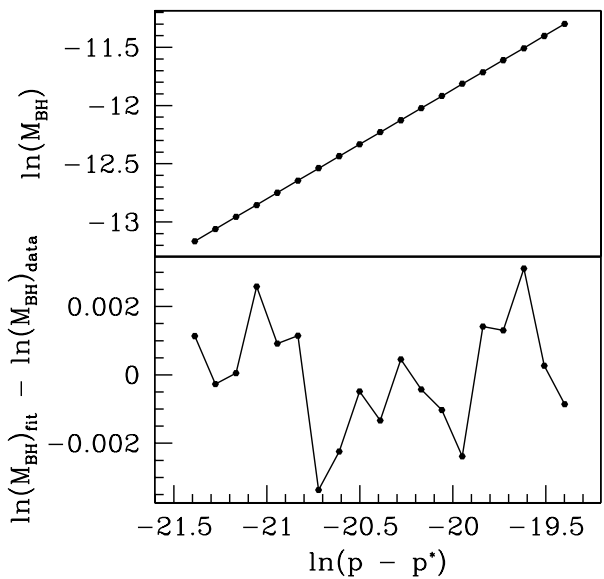}
\caption{Best-fit for the scaling behavior of black hole masses near the threshold.
The top plot shows calculated masses and the fitting line, while the bottom plot shows the 
deviation between the two.  The scaling exponent for this fit, which is simply the slope of 
the line, is $\gamma=0.94$.  
\label{fig:super-scaling-bestfit}}
\end{figure}

To demonstrate the scaling behavior of $M_\mathrm{BH}$, we show in 
Fig.~\ref{fig:super-scaling-alldata}  $\ln\left(M_\mathrm{BH}\right)$ versus 
$\ln|p-p^\star|$ for a wide range of supercritical solutions.  The slope 
of the trend 
is equal to the scaling exponent, $\gamma$.  From the figure, we can clearly
see that the scaling law provides a good fit only in the limit $p\rightarrow p^\star$ as 
expected \cite{koike-etal-1995}. 
The jump seen at $\ln |p-p^\star| \simeq -10$ represents the point at 
which the fluid 
enters a dynamical phase
where the center part of the star has enough kinetic energy to dominate the 
effective potential energy, whose magnitude is 
set by $\rho_\circ$.  In this regime, the fluid then follows a CSS-type evolution.

In addition, Fig.~\ref{fig:super-scaling-alldata}
is meant to illustrate problems in our calculations associated with the coordinate
singularity that inevitably develops in our Schwarzschild-like coordinate system 
in super-critical evolutions. Computations were run for a set of parameter values, $p_i$,
distributed uniformly in $\ln |p-p^\star|$---any gaps in the plotted data thus 
represent instances where our code crashed prematurely.
We note that the flow velocity 
becomes discontinuous and nearly luminal when black hole formation is imminent, and 
 this seems to amplify the instability mentioned 
in Sec.~\ref{sec:instability}. This  results in the evolution halting before 
${\max}_r(2m(r,t)/r)$ exceeds its nominal threshold of 0.995.
However, for a set of parameter values delineated in the figure by dashed lines, 
we were able to find a good fit to a scaling law.  For that subset of data, 
the fit, as well as the data's deviation from the fit,
are shown in Fig.~\ref{fig:super-scaling-bestfit}.  
One measure of how well the black hole masses are described by a relation of the 
form~(\ref{mass-scaling}) is that deviations from the fit are small and apparently
random.  The slope of the trend yields an 
estimated scaling exponent of $\gamma=0.938$. 

\begin{figure}[htb]
\includegraphics[scale=0.4]{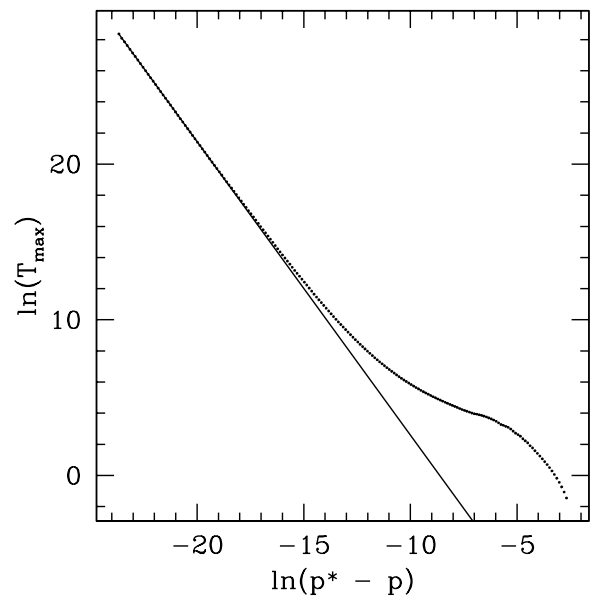}
\caption{Scaling behavior in $T_{\max}$ for subcritical 
solutions, i.e.~those not forming black holes.  
The line shown here is the best-fit for the expected scaling law (\ref{stress-scaling}),
using data only 
from the solutions closest to criticality. 
These runs used $\rho_c = 0.05$, $U = U_1$ and an initial grid defined by
$\{N_a, N_b, N_c, \Delta r_a, \mathrm{level}\} = \{300, 500, 20, 0.005, 0\}$.  
\label{fig:sub-scaling-alldata}}
\end{figure}

\begin{figure}[htb]
\includegraphics[scale=0.4]{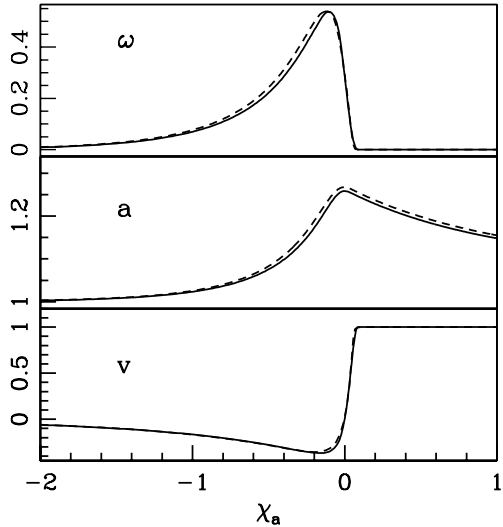}
\caption{Three scale-free quantities of near-critical 
solutions in self-similar coordinates for the ideal-gas system (dashed line) and the ultrarelativistic 
system (solid line). \label{fig:ideal-ultra-critsoln}}
\end{figure}

As mentioned in the previous section,
to obtain another estimate of the scaling exponent, we calculate how the global maximum,
$T_{\max}$, of the stress-energy trace scales as $p \rightarrow p^\star$, 
using subcritical computations.
As Garfinkle and Duncan \cite{garfinkle2} pointed out for the case of 
spherically-symmetric massless scalar 
collapse, the global maximum of the 
Ricci scalar should be proportional to the inverse square of the fundamental length scale of the 
self-similar solution.  Hence $T_{\max}$ for near critical solutions below the threshold should follow 
the scaling law:
\beq{
T_{\max} \propto \left|p - p^\star\right|^{-2\gamma} \, . 
\label{stress-scaling}
}
Using $T$ instead of the Ricci scalar is computationally more expedient  since it does not 
require the calculation of 
second-order space and time derivatives of the metric functions. 

By determining $\gamma$ from a plot of $T_{\max}$ versus $\left|p - p^\star\right|$,
in addition to a fit to Eq.~(\ref{mass-scaling}),  
we can get an estimate of the systematic errors in our estimation of $\gamma$ for both 
methods.
Estimation of $\gamma$ from the scaling of $T_{\max}$ also has the advantage that, 
in the limit $p\to p^\star$, the code is more stable for subcritical, rather than supercritical,
evolutions. 
The scaling behavior for $T_{\max}$ can be seen in  
Fig.~\ref{fig:sub-scaling-alldata} where $\ln T_{\max}$ is plotted versus 
$\ln|p-p^\star|$.
The solutions far from criticality seem to smoothly asymptote toward the critical regime.  
The line shown in this plot uses only those points in the regime that provide the best 
linear fit; a closer view of the points used in the fit are shown, for instance, in 
Fig.~\ref{fig:scaling-difffloors}.  
Since the slope of the line now represents $-2\gamma$ (\ref{stress-scaling}), we find 
from this fit that $\gamma=0.94$, which agrees with 
the value found from the scaling of $M_\mathrm{BH}(p)$ to 
within the estimated systematic error in our computations.

Although our calculated scaling exponents match well to results previously obtained for the 
ultrarelativistic fluid with $\Gamma=2$, 
this does not necessarily say how well the ideal-gas critical solutions compare to the ultrarelativistic 
ones in detail.  To obtain the ultrarelativistic critical solutions, we let an adjustable
distribution of ultrarelativistic fluid free-fall and implode at the origin; specifically, the initial 
data for the fluid 
 is set so that $\tau(r,0)$ is a Gaussian distribution and $S(r,0)=0$, and 
the amplitude of the Gaussian is used as the tuning parameter.  The scale-free functions 
from the critical solutions of the velocity-induced neutron star system and the 
ultrarelativistic system are shown in Fig.~\ref{fig:ideal-ultra-critsoln}.  
Here, the quantity $\omega\equiv\omega(r,t)$ is another 
scale-free function determined from metric and fluid quantities via
\beq{
\omega \ \equiv \ 4 \pi r^2 a^2 \rho  \, .
\label{omega-css}
}
In order to make the comparison between the two solutions, the grid functions were transformed 
into self-similar coordinates $\mathcal{T}$ and $\mathcal{X}$:
\beq{
\mathcal{T} \equiv \ln \left(T^\star_0 - T_0 \right) \, ,
\label{T-ss}
}
\beq{
\mathcal{X} = \ln \left( \frac{r}{r_s} \right) \, .
\label{X-ss}
}
where $T_0$ is the elapsed central proper time,
\beq{
T_0(t) \equiv \int_0^t \alpha(0,t^\prime) \, dt^\prime \, ,
\label{central-propertime}
}
$T^\star_0$ is the accumulation time of a given critical solution, and $r_s(t)$ is the 
location of the sonic point. Since we find that computing a smooth $r_s(t)$ near the 
critical point is difficult, we typically use 
$\mathcal{X}_a$, 
\beq{
\mathcal{X}_a = \ln \left( \frac{r}{r_{a_{\max}}} \right) \, ,
\label{X-ss2}
}
as our self-similar radial coordinate.  Here, $r_{a_{\max}}(t)$ is the position of the local maximum of $a(r,t)$ that lies closest to $r=0$.  

\begin{figure}[htb]
\includegraphics[scale=0.4]{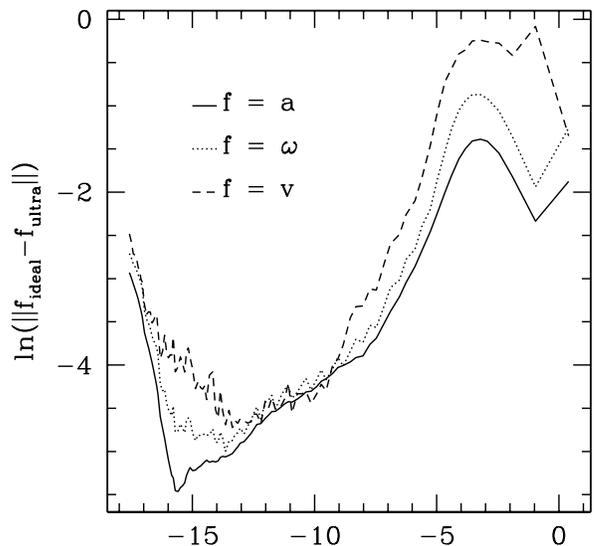}
\caption{Deviation over time of those quantities displayed 
in Fig.~\ref{fig:ideal-ultra-critsoln}.
Here, $||f||$ denotes the $\ell_2$-norm of the function 
$f$.
The deviations for $a$ (solid line), $\omega$ (dotted line), 
and $v$ (dashed line) are shown.  The $\ell_2$-norms of these differences
are computed at every time satisfying $\mathcal{X}_a < 2$, 
and then logarithms of those norms are plotted as a function of self-similar time
 $\mathcal{T}$.  Note that the sense of physical time is opposite to that of 
$\mathcal{T}$; that is, $\mathcal{T}\rightarrow-\infty$ as the solution approaches the 
accumulation time.  As the evolution proceeds from the initial time, the two solutions asymptote 
toward each other.  For $\mathcal{T} < -13$, the deviation between the two solutions increases 
as the ideal-gas near-critical solution departs from the asymptotic critical solution and 
eventually disperses from the origin.  
\label{fig:ideal-ultra-deviation}}
\end{figure}

Our results indicate that the ideal-gas system {\em does} 
asymptote to the ultrarelativistic self-similar solution 
in the critical limit.
While the ultrarelativistic fluid  enters a self-similar phase shortly after the initial time, 
the ideal-gas solution generally tends toward the critical solution 
relatively slowly, then eventually diverges  from 
it.  The agreement between the ideal-gas and ultrarelativistic solutions improves as $p\rightarrow p^\star$, 
as expected, and Fig.~\ref{fig:ideal-ultra-critsoln} shows profiles at a time when the difference 
between the solutions was minimized. The 
$\ell_2$-norms~\footnote{The $\ell_2$-norm of a vector u is defined
as $||u|| = \sqrt{\sum_{i=1}^N u_i^2/N}$.}
of the deviations between the ideal-gas and ultrarelativistic 
scale-free functions plotted over time 
are shown in Fig.~\ref{fig:ideal-ultra-deviation}; it can be easily gleaned from this figure that the 
minimum of the average deviations occurs at  approximately $\mathcal{T}=-13.1$, which is the time at which 
we  have displayed the profiles in Fig.~\ref{fig:ideal-ultra-critsoln}.  
Also, Fig.~\ref{fig:ideal-ultra-deviation}
graphically illustrates how the ideal-gas solution 
asymptotes exponentially---in central proper time, $T$---to the ultrarelativistic critical 
solution 
at early times.  The deviations for the three functions seem to have the same qualitative trend, indicating 
that metric \emph{and} fluid quantities asymptote to their ultrarelativistic counterparts.

\begin{figure}[htb]
\includegraphics[scale=0.4]{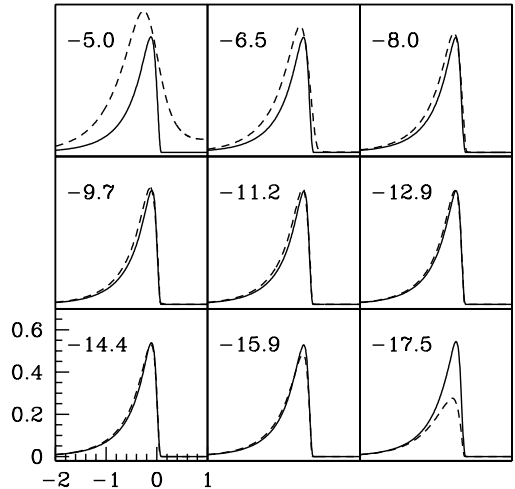}
\caption{Time sequences of $\omega$ for the most nearly critical solutions obtained 
with the ideal-gas EOS (dashed line) and the ultrarelativistic EOS (solid line).  
Both functions have been transformed into self-similar coordinates, based upon 
their respective accumulation times and respective values of $r_{a_{\max}}$.
Approximate values of $\mathcal{T}$ are shown in the upper-left corners of the frames.  
Note that the ultrarelativistic $\omega$ is varying 
slightly frame-to-frame, contrary to appearances.  
Relative to the intrinsically ultrarelativistic solution, it takes more time for the
ideal-gas solution to 
become self-similar since the length scale set by $\rho_\circ$ in the latter 
case only becomes insignificant
for $P/\rho_\circ \gg 1$.  
\label{fig:crit-omega-evolution}}
\end{figure}

This exponential approach of the ideal-gas solution to the self-similar solution 
is better 
seen in the comparison of time sequences of $\omega$ extracted
from the ideal-gas and ultrarelativistic computations, as shown in 
Fig.~\ref{fig:crit-omega-evolution}.  Here, $\omega_\mathrm{ultra}$ 
has already attained a self-similar form at the beginning of the displayed
sequence, while $\omega_\mathrm{ideal}$ becomes self-similar at later times, 
and remains self-similar only for a time interval $\Delta \mathcal{T} \simeq 6$. 

%%%%%%%%%%%%%%%%%%%%%%%%%%
\subsection{Universality and Consistency}
\label{sec:universality-consistency}

This section describes several numerical experiments primarily designed to ensure 
that the results presented above are not 
artifacts of the computational techniques used. 
These computations also provide a measure of the systematic
error in our calculation of $\gamma$.  
Moreover, in order to check previous claims that critical solutions 
generated from perfect fluid configurations having the same adiabatic 
index $\Gamma$ may not reside in the same universality class, we
also measure $\gamma$ for different initial conditions, while keeping $\Gamma$ constant. 
When making any comparisons, the methods, parameters, and initial data used to produce
Figs.~\ref{fig:super-scaling-alldata}--\ref{fig:sub-scaling-alldata} will be referred to as 
the ``original'' configuration.  A tabulation of the values of $\gamma$ and $p^\star$ calculated 
from the different simulation configurations discussed in this section
is given in Table~\ref{table:gammas}. 

\begin{figure}[htb]
\includegraphics[scale=0.4]{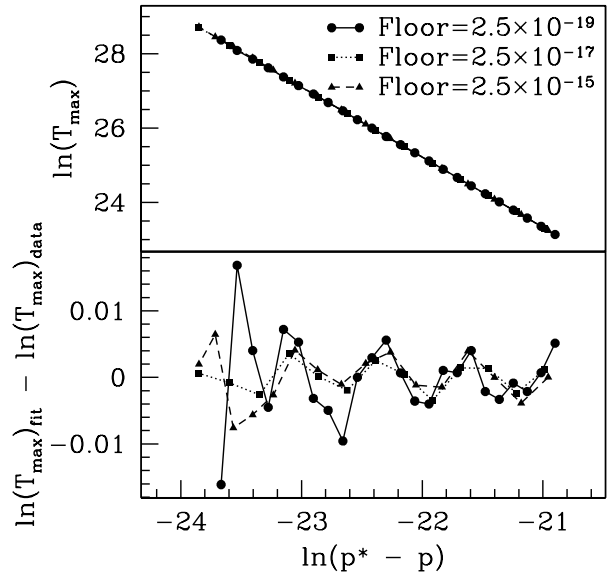}
\caption{Scaling behavior in $T_{\max}$ near the critical 
solution for runs using different values of $\delta$.  
The scaling behaviors are shown for the original configuration (circles,solid line), a configuration with 
$10^2$ times the original floor value (squares, dotted line), and one with 
$10^4$ times the original floor value (triangles, dashed line).
The scaling exponents $\gamma$ for these runs are listed in Table~\ref{table:gammas}
\label{fig:scaling-difffloors}}
\end{figure}

The effect on the scaling behavior due to the 
floor value, $\delta$, used for the fluid (see App.~\ref{app:fluid-methods}) is
estimated first.  Since the magnitude of the floor is set in an
{\em ad hoc} fashion---i.e.~without any physical basis---it is crucial to verify 
that any results are independent of it.
To test this, we replicated the original runs using different values 
of the floor while keeping all other parameters fixed.  
The scaling behavior obtained  using different floor values is illustrated 
in Fig.~\ref{fig:scaling-difffloors}.  The dotted and dashed lines correspond to floor values that 
are factors of $10^2$ and $10^4$, respectively, larger than the original configuration, which itself
used $\delta=2.5\times 10^{-19}$.  The minimal influence of the floor 
on solutions in the critical regime is clearly seen by the fact that all points follow 
nearly the same best-fit line.  In fact, Table~\ref{table:gammas} indicates that all estimated values of 
$\gamma$ agree to within $\simeq0.5\%$ and that all estimates of $p^\star$ 
coincide to within $0.0005\%$.  The deviations of the calculated sets 
$\{\ln\left(T_{\max}\right),\ln|p-p^\star|\}$ from their respective best-fit lines 
for the different floor values even follow the same functional form, suggesting that the 
observed ``periodic'' deviations from linearity are not due to the floor.  

The absence of any dependence on the floor is not too surprising since the component of the fluid that 
undergoes self-similar collapse is never rarefied enough to trigger 
use of the floor.
For instance, at a time when the central part of 
the star begins to resemble an ultrarelativistic critical solution, 
the minimum values of $\{D,\Pi,\Phi\}$ within the sonic point are, respectively, 
$\{\sim10^2, \sim10^3, \sim10^3 \}$---far above the typical floor values used.  Only 
for $r \gtrsim R_\star$ is the floor activated, and dynamics 
in this region cannot affect the interior solution once self-similar collapse 
begins due to the characteristic structure of near-critical solutions 
(see Table~\ref{table:ultrarel-char-speeds}). 

\begin{figure}[htb]
\includegraphics[scale=0.4]{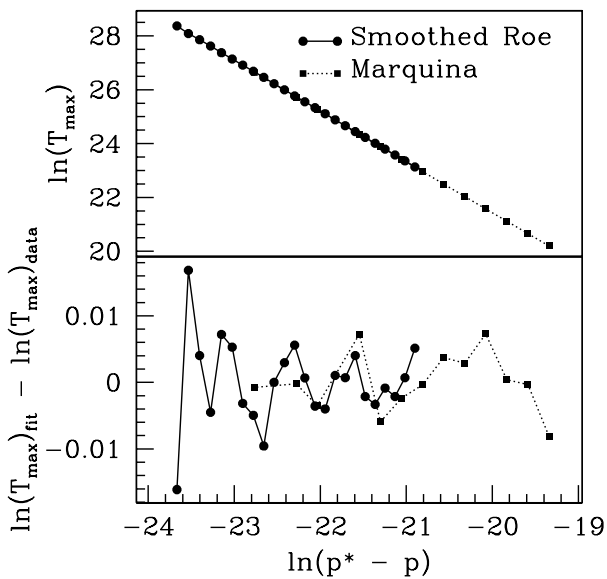}
\caption{Comparison of the scaling behavior in $T_{\max}$ obtained with two different Riemann solvers.  
The ``Smoothed Roe'' line corresponds to the original calculations. 
The other (dotted) line was generated using the Marquina method, with other computational
methods and parameters identical to the original calculations.
The scaling exponents, $\gamma$, for these runs are listed in Table~\ref{table:gammas}
\label{fig:scaling-diffsolvers}}
\end{figure}

The effect from the Riemann solver used on the scaling behavior is seen in Fig.~\ref{fig:scaling-diffsolvers}.
We find that the scaling behavior of $T_{\max}$ from the two methods is remarkably close.  Even though the Roe method 
with smoothing allows us to determine $\ln\left(T_{\max}\right)$ for smaller 
values of $\ln|p-p^\star|$, the deviations from the best-fit  of the 
two data sets are of the same order of magnitude for common values of 
$\ln|p-p^\star|$.
From Table~\ref{table:gammas}, we see that the respective values of $\gamma$ agree to within $0.3\%$ 
and that values of $p^\star$ agree to within $0.001\%$.  These differences are quite small---comparable 
to those found as a result of varying the floor.  Hence, we conclude that the choice in Riemann solvers has 
little, if any, effect on the computed scaling behavior, indicating that the smoothed approximate Roe solver 
is adequate for our purposes.  

\begin{figure}[htb]
\includegraphics[scale=0.4]{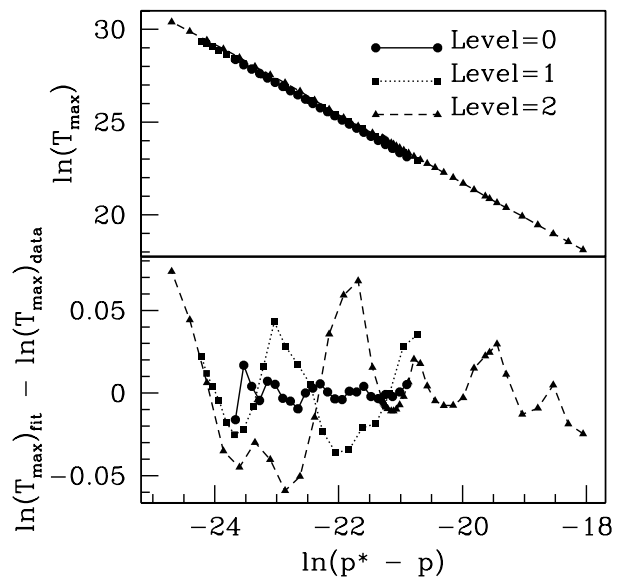}
\caption{Scaling behavior in $T_{\max}$ near the critical 
solution for runs using different levels of resolution.  The runs were made with 
$\rho_c = 0.05$, $U = U_1$, and the solid line with circles was generated from runs using the original configuration.  
The $\mathrm{level}=1$ (squares, dotted line) and $\mathrm{level}=2$ (triangles, dashed line) runs, respectively, used 
computational grids that were locally $2$ times and $4$ times as refined.
The scaling exponents, $\gamma$, for these runs are listed in Table~\ref{table:gammas}
\label{fig:scaling-difflevels}}
\end{figure}

When using finite difference methods, it is vital to 
verify that the order of the solution error is the same as the order to 
which the derivatives are approximated by difference operators.
For example, our HRSC scheme should be $O(\Delta r^2)$
accurate in smooth regions and $O(\Delta r)$ near shocks, so we should expect this scaling 
behavior of the error as $\Delta r$ is varied.  First, we wish to see if our estimate for $\gamma$
converges as the grid is refined.  Figure~\ref{fig:scaling-difflevels} shows a plot of 
$\ln\left(T_{\max}\right)$ versus 
$\ln|p-p^\star|$
for the original configuration,
along with others computed at higher resolutions.  We first see that the three distributions 
follow lines of approximately the same slope (and which are thus shifted vertically
relative to one another by constant amounts) while the deviation of the best-fits
seems to increase slightly with resolution.  Also, we can see that an increase in resolution 
permits us to follow the collapse through to dispersal for solutions closer to the critical threshold,
allowing for the scaling law to be sampled at smaller $\ln|p-p^\star|$.
Even though the deviations from the best-fits for $l=1,2$ are
quite small compared to the typical size of  $\ln\left(T_{\max}\right)$, it is a little worrisome
that they are larger than those from the lowest resolution runs.  However, this
behavior can likely be attributed to the sonic point instability and the smoothing procedure used to 
dampen it.  In particular, the ``bump'' at $r_s$ sharpens with increasing resolution spanning a 
roughly constant number of grid cells (see Sec.~\ref{sec:instability} for more details).  
Consequently, the impact of the instability on the solution may also increase  with 
decreasing $\Delta r$, since the discretized difference operators will generate 
increasingly large estimates for spatial derivatives in the vicinity of the sonic point.
In addition, the smoothing operation is always 
performed using nearest-neighbors, so the smoothing radius physically shrinks with resolution,
diminishing the impact of the smoothing.

\begin{figure}[htb]
\includegraphics[scale=0.4]{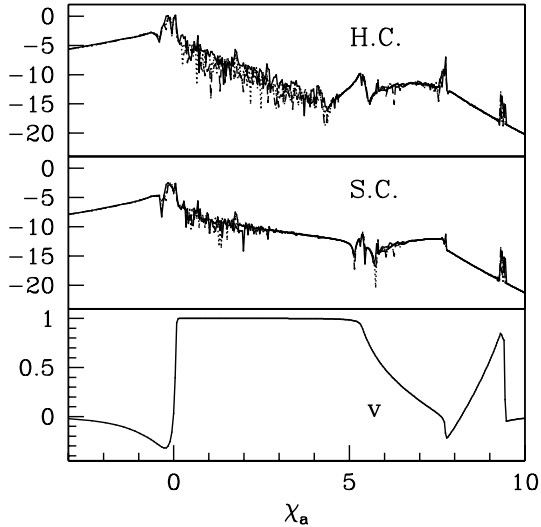}
\caption{Logarithm of scaled, independent 
residuals of the Hamiltonian constraint (\ref{polar-areal-hamiltonian-const}) 
and slicing condition (\ref{polar-areal-slicing-condition}) for three levels of resolutions 
calculated from solutions in the self-similar regime.  The dotted (dashed) lines are from 
a run which used $2$ ($4$) times the local spatial and temporal resolutions of the original run, which is represented 
by the solid lines; 
the dotted (dashed) residual was scaled by a factor of $4$ ($16$) in order to make 
the $O(\Delta r^2)$ convergence of the solution more apparent.
Each distribution is from a solution that has been tuned to  $\ln |p-p^\star| \simeq -19$
with respect to the value of $p^\star$ for each resolution, and function values at every tenth 
grid point are shown.
The physical velocity of the fluid for 
the $l=0$ run is shown in the bottom frame in order to facilitate comparison of features in the 
independent residuals
to those in the solution.  
\label{fig:crit-resids}}
\end{figure}

In order to verify that the code is converging in the self-similar regime, we computed 
independent residuals (i.e.~applied discretizations distinct from those used in
the scheme used to compute the solution) of the Hamiltonian constraint (\ref{polar-areal-hamiltonian-const}) and slicing condition
(\ref{polar-areal-slicing-condition}) for the three levels of resolution 
discussed previously
(Fig.~\ref{fig:crit-resids}).  
The overlap of the scaled residuals seen in the figure indicates $O(\Delta r^2)$ 
convergence.   Note that the smoothing procedure has not been used to calculate the solutions shown here. 
We see that the scaled residuals have similar magnitudes in all regions 
except those that have been processed by shocks, namely $\mathcal{X}_a=[0,4.5],\simeq7.8,\simeq9.4$.
Because the self-similar solutions are converging at the expected rate, we surmise that the 
variations observed in $\gamma$ for the three resolutions does not indicate a problem with convergence, but
demonstrates the effect of truncation error and/or the smoothing procedure on the scaling behavior.  With only 
three levels of resolution, it is hard to make definite claims as to whether $\gamma$ is or is not converging to a 
particular value.  Even so, the standard deviation of $\gamma$ determined from the three evolutions 
is about $1\%$ of their mean, suggesting that the variation is not significant.
In fact, it is comparable to the $2\%$ standard deviation
found from the simpler ultrarelativistic perfect fluid studies of~\cite{neilsen-crit}.

\begin{figure}[htb]
\includegraphics[scale=0.4]{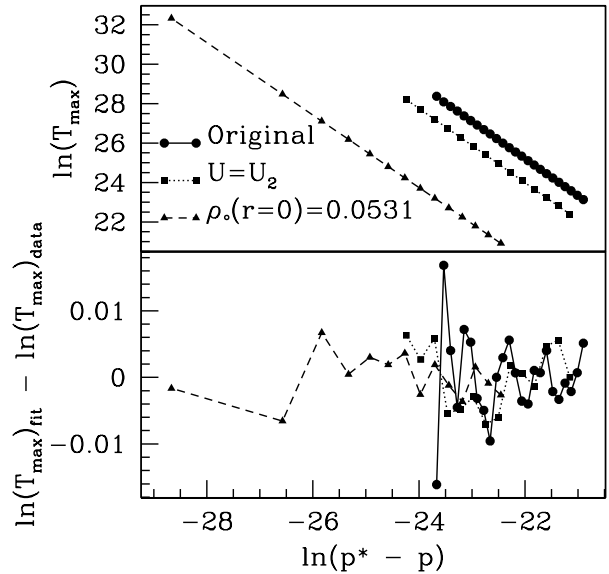}
\caption{Scaling behavior in $T_{\max}$ for several families of initial data.  
The ``Original'' date (solid line) is as before, the dotted line with squares shows the scaling behavior 
for runs that used a different  initial velocity profile, $U = U_2$, and the dashed line with triangles 
was made from  runs with a different TOV solution, with $\rho_c = 0.0531$.  The 
scaling exponents, $\gamma$, for these runs are listed in Table~\ref{table:gammas}
\label{fig:scaling-difftypes}}
\end{figure}

The final comparison entails varying the physical initial conditions of the system 
to investigate the universality of the critical phenomena computed with the ideal-gas 
EOS.  The primary constituents of our model are the initial
star solution and the form of the perturbation with which we drive the star to collapse.  Hence, 
we choose to perform sets of runs to measure the scaling law using 1) a different initial star 
solution and 2) a different functional form of the initial velocity profile.  The scaling behaviors
of $\ln\left(T_{\max}\right)$ versus $\ln|p-p^\star|$ for these different
configurations are compared to the results from the original configuration in 
Fig.~\ref{fig:scaling-difftypes}.  For the data computed using a star 
of central density $\rho_c=0.0531$, the only different aspect is the initial star solution.
This particular value of $\rho_c$ is chosen since it is 
near the transition from Type~II to Type~I phenomena discussed in 
detail in~\cite{noble-choptuik2}.
The second initial data set
uses $U_2$ (\ref{v-profile-12}) for the initial profile of the coordinate velocity.   
Naturally, we see that the three 
data sets are vertically shifted relative to one another since each set 
evolved from significantly
different
profiles of mass-energy---the details of the initial data set the 
scale for $T_{\max}$ for specific values of $\ln |p- p^\star$.
However, only the slopes of the curves are relevant for estimating $\gamma$.

From the values listed in Table~\ref{table:gammas}, we see that $\gamma$ varies more significantly 
with the particular star solution used than with the form of the velocity profile.  In fact, 
we are able to tune closer to the critical solution with the more compact star, a trend 
that can also be seen in 
Table~\ref{table:type-ii-stars}.  
Nonetheless, the scaling exponents computed using the two distinct initial star 
configurations agree to within $2\%$.  

The change in the initial velocity profile only affects the computed value 
of $\gamma$ by $0.04\%$.  
This suggests that other methods of perturbation would also yield close to the same value.
The concordance of results from these three different families of initial data 
imply that universality of critical solutions 
is maintained for perfect fluids governed by an ideal-gas EOS, at least for the 
case $\Gamma=2$.
It would be 
interesting to see whether these properties hold with even more realistic equations of state, as well as for other values of $\Gamma$.

%%%%%%%%%%%%%%%%%%%%%%%%%
\subsection{Final Determination of $\gamma$}
\label{sec:final-determ-gamma}

Using the calculated values of $\gamma$ from the various methods, floor sizes, and
grid resolutions, we are able to provide an estimate of the systematic error inherent in our 
numerical model.  Further, by assuming that the universality is strictly true, we can even use the 
variation in $\gamma$ computed 
from the different initial data families for this estimate.   
Taking the average and calculating the 
standard deviation
from all of the values for the ideal-gas EOS listed in Table~\ref{table:gammas}, 
we estimate a scaling exponent value of 
\beq{
\gamma \ = \ 0.94 \pm 0.01   \, . 
\label{final-gamma}
}
This is in agreement with the value of $\gamma$ computed 
from the black hole mass scaling fit shown in Fig.~\ref{fig:super-scaling-bestfit}. 

\begin{table}
\caption{Scaling exponents $\gamma$ 
and critical parameters $p^\star$ computed from fits to the expected scaling behavior in $T_{\max}$.  
Scaling exponents were obtained from runs using 
initial star solutions with different central densities ($\rho_c$) 
and using different floor magnitudes ($\delta$), levels of refinement ($l$), 
and velocity profiles ($U$).
The runs labeled ``Roe'' use the approximate Roe solver with smoothing, the ``Marquina'' run used 
the Marquina flux formula, and the ``Ultra-rel.'' scaling exponent was computed from our results
involving the collapse of Gaussian profiles of ultrarelativistic fluid.
}
\label{table:gammas}
\begin{ruledtabular}
\begin{tabular}{cdccccc}
Method &\rho_c   &$\delta$ &$l$  &$U$  &$\gamma$ &$p^\star$  \\
\hline
Roe &0.05   &$2.5\times10^{-19}$  &0 &$U_1$ &0.94         &0.4687537   \\ 
Roe &0.05   &$2.5\times10^{-17}$  &0 &$U_1$ &0.94         &0.4687535   \\ 
Roe &0.05   &$2.5\times10^{-15}$  &0 &$U_1$ &0.95         &0.4687516   \\ 
Roe &0.05   &$2.5\times10^{-19}$  &1 &$U_1$ &0.92         &0.4682903    \\ 
Roe &0.05   &$2.5\times10^{-19}$  &2 &$U_1$ &0.93         &0.4682461    \\ 
Roe &0.05   &$2.5\times10^{-19}$  &0 &$U_2$ &0.94         &0.4299032    \\
Roe &0.0531 &$2.5\times10^{-19}$  &0 &$U_1$ &0.92         &0.4482047   \\ 
Marquina  &0.05   &$2.5\times10^{-19}$  &0 &$U_1$ &0.94   &0.4687682    \\ 
%
%Roe &0.05   &$2.5\times10^{-19}$  &0 &$U_1$ &0.94         &0.4687536738   \\ 
%Roe &0.05   &$2.5\times10^{-17}$  &0 &$U_1$ &0.94         &0.4687535028   \\ 
%Roe &0.05   &$2.5\times10^{-15}$  &0 &$U_1$ &0.95         &0.4687516089   \\ 
%Roe &0.05   &$2.5\times10^{-19}$  &1 &$U_1$ &0.92         &0.4682903094    \\ 
%Roe &0.05   &$2.5\times10^{-19}$  &2 &$U_1$ &0.93         &0.4682461196    \\ 
%Roe &0.05   &$2.5\times10^{-19}$  &0 &$U_2$ &0.94         &0.4299031509    \\
%Roe &0.0531 &$2.5\times10^{-19}$  &0 &$U_1$ &0.92         &0.44820474298   \\ 
%Marquina  &0.05   &$2.5\times10^{-19}$  &0 &$U_1$ &0.94   &0.46876822118    \\ 
\hline
Ultra-rel. &-&-&-&-&0.97 &-
\end{tabular}
\end{ruledtabular}
\end{table}

In addition, we can compare our final estimate of $\gamma$ to values previously found for the ultrarelativistic
fluid.  As already mentioned, $\gamma$ 
was measured at three 
different refinement levels in~\cite{neilsen-crit}, and a value
\beq{
\gamma \lesssim  \ 0.96 \, . \label{gamma-neilsen-choptuik}
}
was quoted.

Instead of solving the full set of PDEs, $\gamma$ can also be found by solving the 
eigenvalue problem that results from performing first order perturbation theory about the CSS solution.  
This was done in two ways in \cite{brady_etal}: using the common shooting  method, and solving the 
eigenvalue problem
directly after differencing the equations to second order.   The scaling exponents
calculated were, respectively, $\gamma=0.9386\pm0.0005$ and $\gamma=0.95\pm0.01$. 

%%%%%%%%%%%%%%%%%%%%%%%%%
\subsection{Possible Presence of a Kink Instability}
\label{sec:kink-instability}

As mentioned in Sec.~\ref{sec:instability}, we witness an instability near the sonic point of solutions
tuned close to the threshold.  After thorough numerical experimentation, we are still left uncertain 
about its genesis.  
One possibility is that it has physical origin.  For instance, Harada \cite{harada-2001}
reported the presence of an unstable kink mode for spherically symmetric ultrarelativistic perfect fluids 
when $\Gamma\gtrsim 1.889$.  The unstable kink mode manifests itself as an ever-steepening discontinuity
in $\rho$ at the sonic point, $r_s$, that diverges in finite proper time.  A mild, seed discontinuity at $r_s$ is 
necessary for the mode to grow, but---since it diverges in finite time---minute discontinuities inevitable in 
discretized solutions, for instance, are expected to be sufficient
to manifest the instability.  Physically, the 
``seed'' discontinuity could be the scale at which the continuum approximation of hydrodynamics 
breaks down, i.e.~at the particle scale.  

In our nonlinear PDE solutions that use the ideal-gas EOS, 
we do in fact find a growing discontinuity in $P$ and $\rho_\circ$ 
at $r_s$ (Fig.~\ref{fig:p-rho-ultra-regime}), and this is precisely where our 
instability develops. 

However, as was the case in~\cite{neilsen-crit}, our solutions of the PDEs for an 
explicitly ultrarelativistic fluid do {\em not} develop instabilities in the vicinity of 
$r_s$ for $\Gamma=2$.
One possible reason for this could be the fact that in this case the 
flow is never transonic, i.e.~there is no sonic point since $c_s=1$, and the 
flow can never attain this velocity.   For the ideal-gas fluid, there 
{\em is} a sonic point, as $c_s < 1$ (albeit
arbitrarily close to $1$) since $\rho_\circ/P$ will always be non-zero during a numerical evolution.  
However, the results of~\cite{neilsen-crit} include 
ultrarelativistic near-threshold solutions for $\Gamma=1.99$.
This case {\em does} allow for the presence of the sonic point, so it may be that 
sonic-point kinks were 
not seen in this instance because $\rho$ never became sufficiently steep 
at $r_s$ to excite the kink mode.

\begin{figure}[htb]
\includegraphics[scale=0.4]{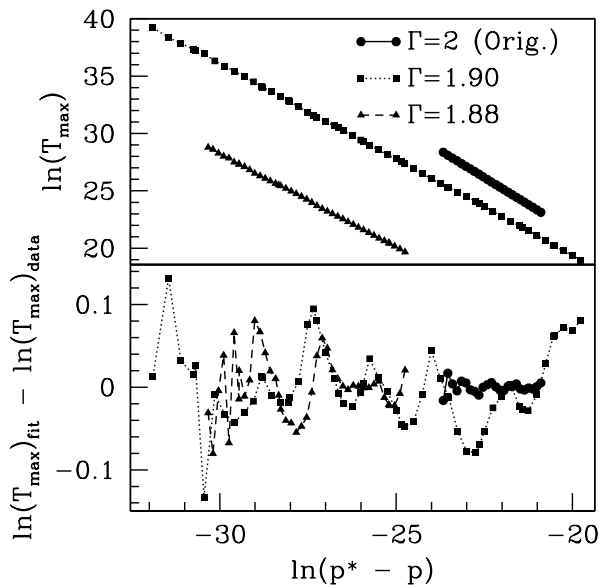}
\caption{Scaling behavior in $T_{\max}$ for different values of $\Gamma$.  
The ``Original'' (solid line with circles) was made from runs with 
$\rho_c = 0.05$ and $\Gamma=2$ as before.   
Both the $\Gamma=1.88$ (dashed line with triangles) and $\Gamma=1.90$ (dotted line with squares) 
runs used $\rho_c=0.013$ with $\{N_a, N_b, N_c, \Delta r_a\} = \{420, 700, 20, 0.005\}$.   
The scaling exponents are given in the text.
\label{fig:scaling-diffgammas}}
\end{figure}

In Fig.~\ref{fig:scaling-diffgammas}, we plot fits of the Type~II scaling behavior of stars with three
different values of $\Gamma$.  The $\Gamma=2$ distribution is our fiducial system, while the other 
two---$\Gamma=1.88,1.90$---use a different progenitor TOV solution having $\rho_c = 0.013$ instead of $\rho_c=0.05$.
The change in initial state was necessary since use of $\rho_c = 0.05$ for $\Gamma = 1.88$ and 
$1.89$ would not necessarily lead to Type~II behavior
using velocity induced collapse.
In addition, the 
values $1.88$ and $1.89$ bound by a 
difference of $\sim 1 \%$ the critical value, $\Gamma_c\simeq1.889$, above which the kink mode becomes 
unstable  \cite{harada-2001}.  The scaling exponents derived from the two new sets 
of computations are
$\gamma(1.9)=0.83$ and  $\gamma(1.88)=0.81$.  The $\Gamma=1.9$ scaling exponent is the same
as 
that calculated with the ultrarelativistic PDEs in \cite{neilsen-crit}, while the $\Gamma=1.88$ scaling exponent 
is consistent with the value obtained for $\Gamma=1.888$, the  value of $\Gamma$ closest 
to 1.88 for which $\gamma$ was computed in~\cite{neilsen-crit}.

We find that there is no consistent behavior in code stability as the $\Gamma_c\simeq1.889$
value is crossed.  In fact, we find the opposite of the expected behavior: we are able to tune the kink-unstable
($\Gamma=1.90$) data closer to $p^\star$ than the kink-stable ($\Gamma=1.88$) data.  We therefore 
find it unlikely that the kink mode is the cause of our sonic point instability.  

Even if the kink mode is unstable for our EOS, tuning toward the CSS 
solution while in the presence of another unstable mode is not without precedent.   
For example, in a study of the spherically-symmetric general relativistic 
harmonic-map (nonlinear sigma model), Liebling \cite{liebling-1998} 
discovered that by judicious choice of an initial data family, he could tune 
to a critical solution that had been shown to have {\em two} unstable modes.
However, it is unclear what relation this work might have to our current study, since 
the type of initial data that we are studying does not seem to have been chosen in
any particularly special way (i.e.~we suspect that the initial data that we have used
is generic, whereas that used by Liebling to tune to the two-mode-unstable solution
was, by construction, non-generic).  What {\em is} clear is that this issue 
requires further investigation, but we will leave that to future studies.

%%%%%%%%%%%%%%%%%%%%%%%%%%%%%%%%%%%%%%%%%%%%%%%%%%%%%%%%%%%%%%%%%%%%%%%%%%%%%%%%%%%%%
%%%%%%%%%%%%%%%%%%%%%%%%%%%%%%%%%%%%%%%%%%%%%%%%%%%%%%%%%%%%%%%%%%%%%%%%%%%%%%%%%%%%%
%%  CHAPTER  %%%%%%%%%%%%%%%%%%%%%%%%%%%%%%%%%%%%%%%%%%%%%%%%%%%%%%%%%%%%%%%%%%%
%%%%%%%%%%%%%%%%%%%%%%%%%%%%%%%%%%%%%%%%%%%%%%%%%%%%%%%%%%%%%%%%%%%%%%%%%%%%%%%%%%%%%
\section{Conclusion}
\label{sec:concl-future-work}

In this work, we simulated spherically-symmetric relativistic perfect fluid flow in the strong-field regime of 
general relatively.  Specifically, a perfect fluid that admits a length scale, for example one that 
follows a relativistic ideal-gas law,  was used to investigate the dynamics of 
compact, stellar objects.  These  stars were modeled as neutron stars by using a stiff equation 
of state, approximating the behavior of some realistic state equations.  Our models were then used
to study the dynamics of neutron stars so far out of equilibrium that they are driven to gravitational  collapse.

Since these systems entail
highly-relativistic fluid motions and strong, nonlinear effects from the fluid-gravitational 
interaction, a numerical treatment is challenging.  To achieve stable evolutions in
 near-luminal flows, while using high resolution shock capturing techniques, 
the primitive variable solver required improvements.  In addition, an instability was found to 
develop in calculations near the threshold of black hole formation, and this necessitated
the use of
new computational methods, that were only partially successful in stabilizing the 
calculations.

We find our value for the scaling exponent, $\gamma$, 
given in Eq.~(\ref{final-gamma}) agrees well with those found in \cite{brady_etal}, 
and agrees with the value of $\gamma$ computed in ~\cite{neilsen-crit} to within 
the uncertainty quoted in that work.
We note that a discrepancy in the values of $\gamma$ computed using ideal-gas 
and ultrarelativistic fluids was observed in~\cite{neilsen-crit}, and this is also the 
case for our calculations.
Our ultrarelativistic value, 
$\gamma=0.97$, agrees well with the value calculated in~\cite{neilsen-crit}, but 
deviates by an estimated 3 standard deviations from 
the value extracted from our ideal-gas calculations.  It is somewhat interesting, yet probably 
coincidental, that our results from the ideal-gas system of equations lead to estimates of $\gamma$ 
that agree with the perturbation calculations better than those values found from the 
ultrarelativistic PDE calculations.  

Our findings thus do not support some of the results found, and claims made
by Novak \cite{novak} for the case of fluid collapse with an ideal-gas EOS and 
$\Gamma=2$.  This previous work suggested that the Type II behavior observed in such
a case was {\em not} well approximated by a universal (with respect to 
initial data) ultrarelativistic limit.
However, using different stars and  velocity profiles, 
and by varying other aspects of the numerical model, we have found 
scaling behavior that is 
insensitive to approximations made in the numerical solution, and which 
{\em does} appear to be
universal with respect to families of initial data.  
Moreover, we have found that the scaling exponent and critical solution for the collapse 
governed by the ideal-gas EOS
agrees well with their ultrarelativistic counterparts.  

Ultimately, it is our goal to 
expand the model a great deal, making the matter description more realistic and eliminating 
symmetry.  As a first step, we wish to develop adaptive mesh refinement procedures 
for conservative systems that will be  required to study critical phenomena of stellar 
objects in axial-symmetry \cite{choptuik-etal-2003}.  

It remains to be seen whether the universal scaling behavior we have observed is also seen
with more realistic state equations such as the one Novak used.  Since accurate measurements 
of $\gamma$ have only been found for equations of state with constant adiabatic index $\Gamma$, 
and since $\gamma$ seems to only depend on $\Gamma$ for perfect fluids, 
it will be interesting to investigate in detail
what the scaling behavior---if any---will be like for realistic state equations with 
variable $\Gamma$.  However, to the extent that any given realistic EOS admits a unique
ultrarelativistic limit, characterized by a {\em single} value of $\Gamma$, we can expect 
to see universal Type II behavior of the sort discussed in this paper, for at least {\em some}
collapse scenarios.

%%%%%%%%%%%%%%%%%%%%%%%%%%%%%%%%%%%%%%%%%%%%%%%%%%%%%%
%%%%%%%%%%%%%%%%%%%%%%%%%%%%%%%%%%%%%%%%%%%%%%%%%%%%%%
\begin{acknowledgments}

We wish to acknowledge financial support from CIFAR, NSERC, and NSF PHY 02-05155.
SCN wishes to thank I. Olabarrieta for many helpful comments and his ENO code.  
All numerical calculations were performed on UBC's \textbf{vn PIII} and \textbf{vnp4} clusters (supported by 
CFI and BCKDF).

\end{acknowledgments}

%%%%%%%%%%%%%%%%%%%%%%%%%%%%%%%%%%%%%%%%%%%%%%%%%%%%%%%%%%%%%%%%%%%%
%%%%%%%%%%%%%%%%%%%%%%%%%%%%%%%%%%%%%%%%%%%%%%%%%%%%%%%%%%%%%%%%%%%%
%%%%%%%%%%%%%%%%%%%%%%%%%%%%%%%%%%%%%%%%%%%%%%%%%%%%%%%%%%%%%%%%%%%%
%   APPENDIX   %%%%%%%%%%%%%%%%%%%%%%%%%%%%%%%%%%%%%%%%%%%%%%%%%%%%%%
%%%%%%%%%%%%%%%%%%%%%%%%%%%%%%%%%%%%%%%%%%%%%%%%%%%%%%%%%%%%%%%%%%%%
%%%%%%%%%%%%%%%%%%%%%%%%%%%%%%%%%%%%%%%%%%%%%%%%%%%%%%%%%%%%%%%%%%%%
\appendix

%%%%%%%%%%%%%%%%%%%%%%%%%%%%%%%%%%%%%%%%%%%%%%%%%%%%%%%%%%%%%%%%%%%%
%%  APPENDIX A  %%%%%%%%%%%%%%%%%%%%%%%%%%%%%%%%%%%%%%%%%%%%%%%%%%%%%%%%%%%
%%%%%%%%%%%%%%%%%%%%%%%%%%%%%%%%%%%%%%%%%%%%%%%%%%%%%%%%%%%%%%%%%%%%
%%%%%%%%%%%%%%%%%%%%%%%%%%%%%%%%%%%%%%%%%%%%%%%%%%%%%%%%%%%%%%%%%%%%
%%%%%%%%%%%%%%%%%%%%%%%%%%%%%%%%%%%%%%%%%%%%%%%%%%%%%%%%%%%%%%%%%%%%
\section{Finite Difference Equations and Approximate Riemann Solvers}
\label{app:fluid-methods}

We employ High-Resolution Shock Capturing (HRSC) algorithms to solve the equations of 
motion for the fluid (\ref{conservationeq}).  Such methods have become increasingly popular 
in the field of relativistic hydrodynamics since they 
are flux conservative, and ensure that discontinuities are well resolved and propagate 
at the correct speeds in the continuum limit.
A key ingredient to these schemes is their use of solvers 
for the Riemann problem at every cell interface.  This is crucial for the conservative nature of 
these schemes since the solution to the Riemann problem is 
always a weak solution of the hyperbolic conservation laws.  
The ``high-resolution'' aspect of the algorithms denotes that in regions where the grid 
functions are smooth, 
the integration procedure is at least $O(\Delta r^2)$ accurate.
Many of the HRSC methods used in this paper have been used in previous works such as \cite{romero}, 
\cite{novak}, and \cite{neilsen} to name only a few relevant sources.  Also, excellent references 
on conservative methods for general systems of hyperbolic conservation equations
have been written by LeVeque \cite{leveque1,leveque2}.  

Unlike finite difference methods, finite volume or conservative methods  calculate the cell-averages of grid 
functions  instead of the grid functions themselves.  The difference equations for conservative methods 
are not derived from Taylor-series approximations of derivatives, but from differences of integrals.  
For instance, we difference the fluid EOM (\ref{conservationeq}) in the following manner:
\beqa{
\mathbf{\bar{q}}^{n+1}_i \ & = & \ \mathbf{\bar{q}}^{n}_i  \nonumber \\ 
& \, - \, & \frac{3 \, \Delta t}{r^3_{i+1/2} - r^3_{i-1/2}} 
\left[ \left( r^2 X \mathbf{F} \right)^n_{i+1/2}  \right. \nonumber \\ 
& - & \left. \left( r^2 X \mathbf{F} \right)^n_{i-1/2}  \right]  \label{ss-discreteq}  \\ 
& \, + \, & \Delta t \, \bar{\bm{\psi}}^n_i \nonumber 
}
At first glance, this equation seems no different than a finite difference 
approximation of Eq.~(\ref{conservationeq}).
However, the difference between the two approaches becomes apparent when we examine 
how specific quantities in~(\ref{ss-discreteq}) are defined.
$\mathbf{\bar{q}}^{n}_i$ is the spatial average of $\mathbf{q}(r,t)$ over the cell centered at $(r_i,t^n)$, 
$\bar{\bm{\psi}}_i^n$ is the spatio-temporal average of the source function 
$\bm{\psi}$ centered at $(r_i,t^{n+1/2})$, and  the numerical flux $\mathbf{F}_{i+1/2}^{n}$
is the time average of $\mathbf{f}(r_{1+1/2},t)$ from $t^n$ to $t^{n+1}$.  In practice, we approximate 
$\bar{\bm{\psi}}_i^n$ as the source of the averages, $\bm{\psi}(\mathbf{\bar{q}}(r_i,t^{n+1/2}))$. 

The techniques for determining the 
numerical flux are especially important since they set the spatial accuracy of the overall scheme and 
are primarily responsible for how well shocks are resolved by the method.
Unless otherwise stated, all results presented in this paper
were produced using an approximate Roe-type solver as outlined in \cite{romero} for calculating the numerical 
flux.  The Roe solver \cite{roe-1981} approximately solves the Riemann problem at each cell interface by 
casting the conservation equation (\ref{conservationeq}) into quasi-linear form.
We approximate the Roe matrix $\mathbf{A}$ as
\beq{
\mathbf{A}(\mathbf{q}^L,\mathbf{q}^R) 
= \left. \pderiv{\mathbf{f}}{\mathbf{q}}\right|_{\mathbf{q}=\hat{\mathbf{q}}}\ , 
\label{approx-roe-matrix}
}
with
\beq{
\hat{\mathbf{q}} = \frac{1}{2} \left( \mathbf{q}^L + \mathbf{q}^R \right) \, . 
\label{approx-roe-vector}
}
and where $\mathbf{q}^L$ and $\mathbf{q}^R$ will be defined shortly.
After solving the linear Riemann problem, the numerical flux of is computed using the 
following expression~\cite{leveque1}:
\beqa{
\mathbf{F}_{k+1/2}(t) & = & \frac{1}{2} \left[ \mathbf{f}(\mathbf{q}^L_{k+1/2}(t)) + 
\mathbf{f}(\mathbf{q}^R_{k+1/2}(t)) \right. \nonumber \\ 
& - & \left. \sum_m \left| \lambda_m \right| \omega_m \bm{\eta}_m \right]  \, . 
\label{num-flux}
}
Here, $\lambda_m$ and $\bm{\eta}_m$ are the eigenvalues and right eigenvectors, respectively,
of $\mathbf{A}$, $\mathbf{q}^L$ and $\mathbf{q}^R$ are, respectively, the values 
of $\mathbf{q}$ to the left and right of the cell boundary, and $\omega_m$ are the decomposed 
values of the jumps in the space of characteristic values:
\beq{
\mathbf{q}^R - \mathbf{q}^L \ = \ \sum_m \omega_m \bm{\eta}_m  \, . 
\label{decomposed-jumps}
}
In order to calculate all quantities associated with $\mathbf{A}$, such as $\lambda_m$ and 
$\bm{\eta}_m$, we use the average of the left and right states, 
$\hat{\mathbf{q}}$, defined by~(\ref{approx-roe-vector}).

As far as we know, the eigenvectors for our new formulation,
(\ref{conservationeq}) and (\ref{ideal-piphi-state-vectors}), have not previously been 
published, so for completeness we present them here. 
Since the transformation from $\{D, S, \tau\}$ to $\{D, \Pi, \Phi\}$ is linear, the eigenvalues 
remain the same:
\beq{
\lambda_1 = v 
\qquad , \qquad  
\lambda_{2 \atop 3} \ = \ \lambda_\pm \ = \ \frac{ v \pm c_s }{ 1 \pm v c_s }  \, .
\label{ideal-piphi-evalues}
} 
Using \texttt{Maple} and assuming the the ideal-gas EOS (\ref{ideal-eos}), we have calculated 
the left and right eigenvectors; the Marquina flux formula requires the right eigenvectors
\cite{donat-marquina}.   Using the typical normalization for the
eigenvectors ($\bm{\eta}_m^{(2)} = \lambda_m$), leads to a 
very complicated set of eigenvectors.  Hence, we used the normalizations
\beq{
\bm{\eta}_m^{(1)} = 1   \quad \quad  \forall \ m  \, ,
\label{ideal-piphi-evect-normalization}
}
which leads to significant simplification.  
The right eigenvectors become:
\beq{
\bm{\eta}_1 \ = \ \left[ \begin{array}{c} 1 
\\[0.25cm]  \frac{W \left(1+v\right)}{a} - 1 
\\[0.25cm]  \frac{W \left(1-v\right)}{a} - 1 \end{array}\right]   \, , \quad
\label{ideal-piphi-right-evects-1}}
\beq{
\bm{\eta}_{2 \atop 3}  = \bm{\eta}_\pm \ = \ \left[ \begin{array}{c} 1 
\\[0.25cm] \frac{ W  \left( 1 + v \right) }{ a }  h \left( 1 \pm c_s \right) - 1
\\[0.25cm] \frac{ W  \left( 1 - v \right) }{ a }  h \left( 1 \mp c_s \right) - 1
       \end{array}\right]  , 
\label{ideal-piphi-right-evects-23}
}
while the associated left eigenvectors are
\beqa{
\mathbf{l}_1 \ & = & \ \left[ \begin{array}{c} 
  1  + \frac{\tilde{\kappa}}{h \,c_s^2} \left( 1 - a W \right)
\\[0.25cm]  -\frac{a}{2 h \,c_s^2} \, \tilde{\kappa} \, W \left( 1 - v \right)
\\[0.25cm]  -\frac{a}{2 h \,c_s^2} \, \tilde{\kappa} \, W \left( 1 + v \right)
    \end{array}\right]^T \, ,  \label{ideal-piphi-left-evects1} \\[0.3cm]
\mathbf{l}_{2 \atop 3} \ = \ \mathbf{l}_\pm \ = & \frac{1}{2 h \, c_s^2} &
\left[ \begin{array}{c} 
a W \left( \tilde{\kappa} \mp v c_s \right) - \tilde{\kappa} 
\\[0.25cm]  \frac{1}{2} a W \left(1 - v \right) \left( \tilde{\kappa} \pm c_s \right)
\\[0.25cm]  \frac{1}{2} a W \left(1 + v \right) \left( \tilde{\kappa} \mp c_s \right)
    \end{array}\right]^T \, .
\label{ideal-piphi-left-evects2}
}
In the above expressions we have
\beqa{
c_s^2 & = & \frac{ \left( \Gamma - 1 \right) \Gamma P }
                   { \left( \Gamma - 1 \right) \rho_\circ + \Gamma P } \, , \nonumber \\
\tilde{\kappa} & = & \Gamma - 1  \, ,  \label{ideal-eos-quantities} \\ 
h {c_s}^2 & = & \frac{ \Gamma P }{ \rho_\circ }  \, . \nonumber
}

The simple form the numerical flux (\ref{num-flux}) takes in the approximate Roe method allows us 
to subtly alter the equations of motion in a way that greatly improves the regularity of the conserved variables
near the origin \cite{neilsen}.  Let us first note that the flux term in Eq.~(\ref{conservationeq}) can be expanded 
in a manner that yields a new EOM:
\beq{ 
\partial_t \mathbf{q} 
+ \frac{1}{r^2} \partial_r \left( r^2 X \mathbf{f}^{(1)} \right) +  \partial_r \left( X \mathbf{f}^{(2)} \right) 
= \hat{\bm{\psi}} 
\, , \label{new-conservationeq}
}
where $\mathbf{q}$ remains is unchanged from Eq.~(\ref{ideal-piphi-state-vectors}), and 
\beqa{
\mathbf{f}^{(1)} & \equiv & \left[ \begin{array}{c} D v \\ v \left( \Pi + P \right) 
        \\ v \left( \Phi + P \right) \end{array} \right] \, , \nonumber \\
\mathbf{f}^{(2)} & \equiv & \left[ \begin{array}{c} 0 \\  P  \\  -P \end{array} \right] \ , \   \label{new-flux-vectors} \\ 
\hat{\bm{\psi}} & \equiv & \left[ \begin{array}{c} 0 \\  \Theta  \\  -\Theta  \end{array} \right]   \, . \nonumber 
}
This new formulation eliminates the inexact cancellation of the $2PX/r$ terms from the flux and the source that 
would normally arise from truncation error.  Note that the eigensystem, used by Roe's numerical flux function 
(\ref{num-flux}), is still calculated from the total flux function, 
$\mathbf{f} = \mathbf{f}^{(1)} + \mathbf{f}^{(2)}$.

Due to the finite precision of the calculations and the nature of the 
numerical methods employed,  the evacuation of fluid often ``overshoots''
the vacuum state generating negative pressures or densities, which in turn leads
to problems such as a complex $c_s$ or super-luminal
velocities.  In order to alleviate such problems we require the dynamic
fluid quantities---the conserved variables $\mathbf{q}$ 
(\ref{ideal-piphi-state-vectors})---to have values greater than or equal to 
a so-called ``floor'' state.  In order to determine the floor state, we require 
$P,\rho_\circ > 0$ and $|v| < 1$ which implies that 
\beq{
D \ , \ \left( \tau \pm \left|S\right| \right) \  > \  0  \,\, .
\label{floor-requirements}
}
Using the transformed (``new'') variables $\Pi,\Phi$, we implement this requirement in the following way 
\beqa{
D \ & = & \ {\max}\left(D , \delta \right)\, , \\
\Pi \ & = & \ {\max}\left(\Pi + D , 2 \delta \right) - D \, ,\\
\Phi \ & = & \ {\max}\left(\Phi + D , 2 \delta \right) - D\, , 
\label{floor-method}
}
where $\delta$ is the adjustable floor parameter.
Notice that $\Pi$ and $\Phi$ need not remain positive since $\tau \le 0$ is physical as 
long as $E>0$.  

%%%%%%%%%%%%%%%%%%%%%%%%%%%%%%%%%%%%%%%%%%%%%%%%%%%%%%%%%%%%%%%%%%%%
%%%%%%%%%%%%%%%%%%%%%%%%%%%%%%%%%%%%%%%%%%%%%%%%%%%%%%%%%%%%%%%%%%%%
\section{Boundary Conditions}
\label{app:bound-cond}

For the outer boundary condition of our fluid quantities, we use the typical outflow condition 
where the fluid quantities associated with the last physical cell are copied into so-called
ghost cells (i.e.~first order
extrapolation).  Our experience, as well as that of others, indicates that 
this condition is fairly robust and non-reflective so no other methods were tested or used.

The regularity conditions at the origin are, however, more sophisticated.  Since our fluid grid functions are
defined with respect to a grid that is offset from the origin, typical
$O(\Delta r^2)$ regularity conditions are not as well-behaved as they are for 
origin-centered cells.  Hence, we have found it helpful to use higher-order, conservative 
interpolation for the fields on the first physical cell.  Since the fluid fields, 
$\bar{\mathbf{q}}_i$, are to be interpreted as cell-averages of conserved functions,
which we will call $\mathbf{Q}(r)$, an interpolation is said to be conservative if the 
integral of a function on a local domain is conserved by the interpolation procedure.
We first assume that the interpolation function $\mathbf{Q}_i(r)$ that is associated with 
a cell $\mathcal{C}_i$  has a polynomial expansion of degree $N-1$:
\beq{
\mathbf{Q}_i (r) = \sum^{N-1}_{n = 0} \mathbf{a}_n  
\left( r - {r}_i \right)^n \, ,
\label{interp-function-def}
}
with $N$ coefficients $\mathbf{a}_n$.  
These coefficients are found by demanding that $\mathbf{Q}_i$ maintains
conservation locally.  That is, a set $\mathcal{S}_{i}$ of $N$ cells is chosen in the neighborhood 
of cell $\mathcal{C}_i$, and we require that $\mathbf{Q}_i$ reproduces the known values 
$\bar{\mathbf{q}}_k$, where 
$\mathcal{C}_k \in \mathcal{S}_i$.  Specifically, the coefficients 
$\mathbf{a}_n$ are calculated by solving the following set of $N$ equations:
\beqa{
\bar{\mathbf{q}}_k & \ = \ & \frac{1}{V_k} \int_{V_k} \, \mathbf{Q}_i(r)  \ dV \nonumber \\ 
&\ = \ & \frac{3}{ r_{k+1/2}^3 - r_{k-1/2}^3} \label{ci-ss} \\
& \times & \sum^{N-1}_{n = 0} \mathbf{a}_n  
\left[ \int^{r_{k+1/2}}_{r_{k-1/2}}
\left( r - {r}_i \right)^n \ r^2  dr \right]  \, , \nonumber
}
for all $\mathcal{C}_k \in \mathcal{S}_i$.
Since this interpolation procedure is used at the origin where local flatness is demanded, 
we can assume $a(r,t)=1$ for $r\approx 0$ with negligible effect (in principle, $a(r,t)$ should
appear in the above integral as part of the volume element).
Once Eq.~(\ref{ci-ss}) is solved for the coefficients, $a_n$, 
the interpolation procedure is completed by using~(\ref{ci-ss})
to determine values $\qbar_j$ 
for a cell $\mathcal{C}_j \notin \mathcal{S}_i$.

From the demand of regularity at the origin, the fields $\rho_\circ, P, D, \tau$
are all even in $r$ as $r \to 0$, while $v$ and $S$ are odd.   
Thus, $a_n=0$ for odd $n$ in the interpolation function of the even fields, and 
$a_n=0$ for even $n$ in the odd interpolations.  In our case, the cells lying nearest the origin 
are spaced uniformly and we use $N=4$. For even functions the boundary condition
then becomes
\beq{
{\bar{\mathbf{q}}}_{_{1}} \ = \  \frac{
   3311 \, {\bar{\mathbf{q}}}_{_{2}} 
 - \, 2413 \, {\bar{\mathbf{q}}}_{_{3}} 
 + \, 851  \, {\bar{\mathbf{q}}}_{_{4}} 
 - \, 122  \, {\bar{\mathbf{q}}}_{_{5}} }{1627} \,\, ,
}
while for odd functions we have
\beq{
 {\bar{\mathbf{q}}}_{_{1}} \ = \  \frac{
   35819 \, {\bar{\mathbf{q}}}_{_{2}}
 - \, 16777 \, {\bar{\mathbf{q}}}_{_{3}} 
 + \, 4329  \, {\bar{\mathbf{q}}}_{_{4}} 
 - \, 488   \, {\bar{\mathbf{q}}}_{_{5}} }{36883} \,\, .
}
Here, cell $\mathcal{C}_{1}$ is the innermost cell, located at $r_1 = \Delta r / 2$.  
Since $\Pi$ and $\Phi$ are combinations of even and odd functions, their regularity conditions are not 
as straightforward to compute.  
To determine their behavior at the origin, we first calculate
the interpolated values of $\tau$ and $S$ at  $\mathcal{C}_{1}$, since their regularity behavior
is known.  Then, $\Pi$ and $\Phi$ are calculated for $\mathcal{C}_{1}$ from the definitions 
(\ref{Pi-Phi-ideal}) using the interpolated values of $\tau$ and $S$.

In contrast to the fluid variables, the metric grid functions are defined on a grid 
centred on the origin.  This enables us to 
use straightforward discrete expressions to ensure the regularity of $\alpha$ and $a$. 
In solving the Hamiltonian equation (\ref{polar-areal-hamiltonian-const}), we demand that spacetime be 
locally flat at the origin; this implies $a(0,t)=1$.  
The slicing condition (\ref{polar-areal-slicing-condition}) is solved by integrating inward from the 
outer boundary, making use of the freedom we have in relabeling constant $t$ surfaces.  If we assume 
that all the matter  remains within our grid, then the metric
exterior to the grid is a piece of the Schwarzschild solution.  
Since the Schwarzschild metric 
is asymptotically flat, we can rescale $\alpha$ after the solution of the slicing 
equation so that our metric is equivalent to the standard
Schwarzschild form at $r_{\max}$.  Thus, our boundary condition for $\alpha$ is 
\beq{
\alpha(r_{\max}) = \frac{1}{a(r_{\max})} \, .
\label{alpha-bc}
}
The physical meaning of this commonly-adopted parameterization  is that coordinate time 
and proper time coincide as $r \to \infty$.

%%%%%%%%%%%%%%%%%%%%%%%%%%%%%%%%%%%%%%%%%%%%%%%%%%%%%%%%%%%%%%%%%%%%
%%%%%%%%%%%%%%%%%%%%%%%%%%%%%%%%%%%%%%%%%%%%%%%%%%%%%%%%%%%%%%%%%%%%
\section{Numerical Tests}
\label{app:numerical-tests}

Here, we present a series of tests that verify that our code solves the
equations we claim that it solves, and that our discrete solutions 
converge as expected in the continuum limit.

\begin{figure}[htb]
\includegraphics[scale=0.38]{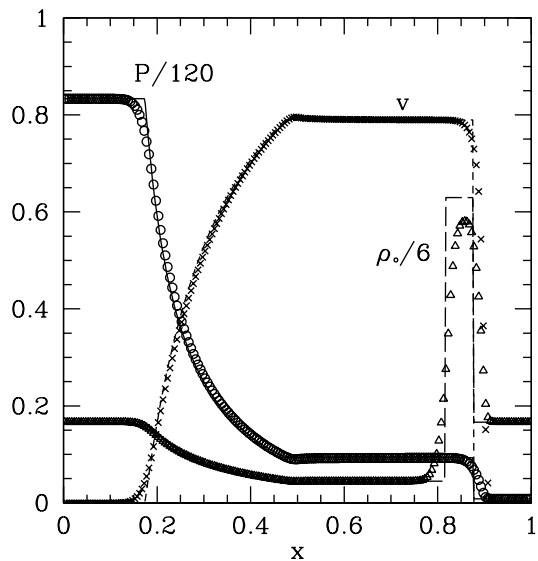}
\caption[Riemann solution using the approximate Roe method 
with initial data 
$\{P^L,v^L,\rho_\circ^L\}=\{100, 0, 1 \}$, $\{P^L,v^L,\rho_\circ^L\}=\{1, 0, 1 \}$ and $\Gamma=5/3$,
using 200 cells.]{Riemann solution using the approximate Roe method 
with initial data $\{P,v,\rho_\circ\}(x<0.5)=\{100, 0, 1 \}$, 
$\{P,v,\rho_\circ\}(x>0.5)=\{1, 0, 1 \}$, $\Gamma=5/3$, using 200 cells.  $P(x)/120$ (circles), 
$\rho_\circ(x)/6$ (triangles), and $v(x)$ ($\times$) are plotted at $t=0.4$. 
The lines correspond to the exact solution.   
\label{fig:roe-riemann}}
\end{figure}

\begin{figure}[htb]
\includegraphics[scale=0.38]{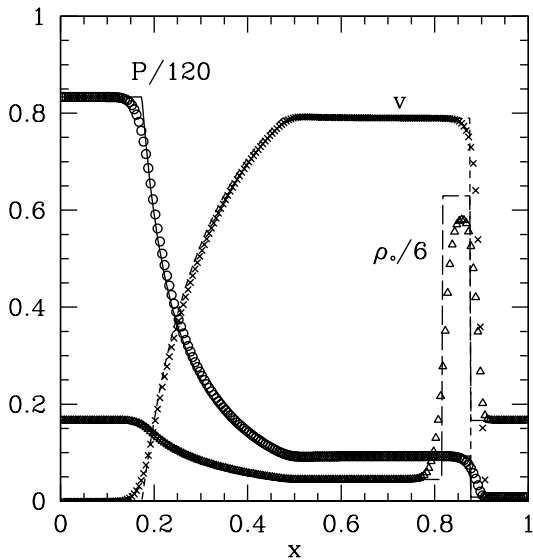}
\caption[Same as in Fig.~\ref{fig:roe-riemann} except using the Marquina method.]{
Same as in Fig.~\ref{fig:roe-riemann} except using the Marquina method.
\label{fig:marq-riemann}}
\end{figure}

In Fig.~\ref{fig:roe-riemann}, we show solutions of a Riemann problem
generated using 1) the approximate Roe solver described previously, and 2) Marquina's method. 
Here, the Riemann problem was set up in the middle 
of the grid, i.e.~at $x=0.5$.  The solid line shows the exact solution of the Riemann problem calculated by 
a routine given in \cite{marti-muller-living}.  The approximate solutions compare favorably to the 
exact solution, especially in smooth regions where the discrete solutions should be close to the 
exact solution.  It seems that Roe's method produces a slight Gibbs phenomenon near 
the origin of the rarefaction fan, while Marquina's method results in undershooting 
in that region.
However, the two methods produce nearly identical results near the shock and contact discontinuity. 

\begin{figure}[htb]
\includegraphics[scale=0.4]{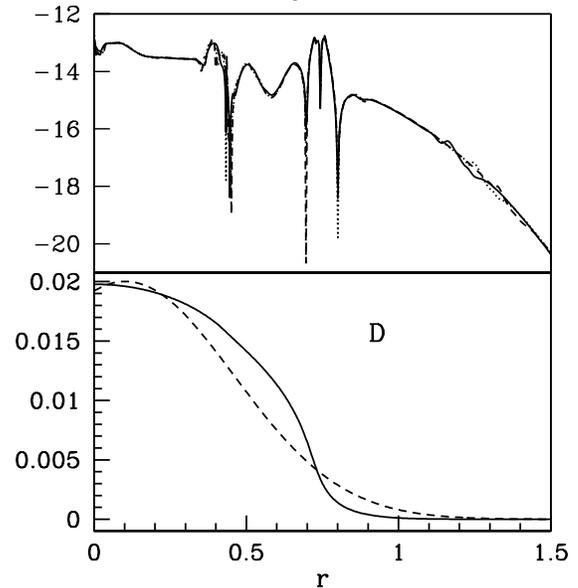}
\caption{Convergence test for the fluid variable $D$.  The top panel shows $\ln\left|D^{8h} - D^{4h}\right|$ 
(solid line), $\ln\left|4\left(D^{4h} - D^{2h}\right)\right|$ (dots), 
$\ln\left|16\left(D^{2h} - D^{h}\right)\right|$ (dashes) which have been scaled so that 
they will coincide if they are well-described by Richardson expansions of the 
form~(\ref{richardson}).  Quantities
with superscript ``$l h$'' have been calculated using grid point separations $l$ times larger than the 
fiducial $l=1$ quantities.
The bottom panel shows  $D(r,0)$ (dashes) and $D(r,t)$ (solid line), where $t$ is the time at which we performed 
the convergence test.  The initial data consisted of a self-gravitating fluid specified by a Gaussian function for 
$\rho_\circ$ centered at $r=0.1$ with an initial
linear velocity profile. The initial grid used for the coarsest solution shown is defined by the parameters 
$\{N_a,N_b,N_c,\Delta r_a\}=\{ 200, 300, 20, 0.005 \}$.
\label{fig:D-converge}}
\end{figure}

\begin{figure}[htb]
\includegraphics[scale=0.38]{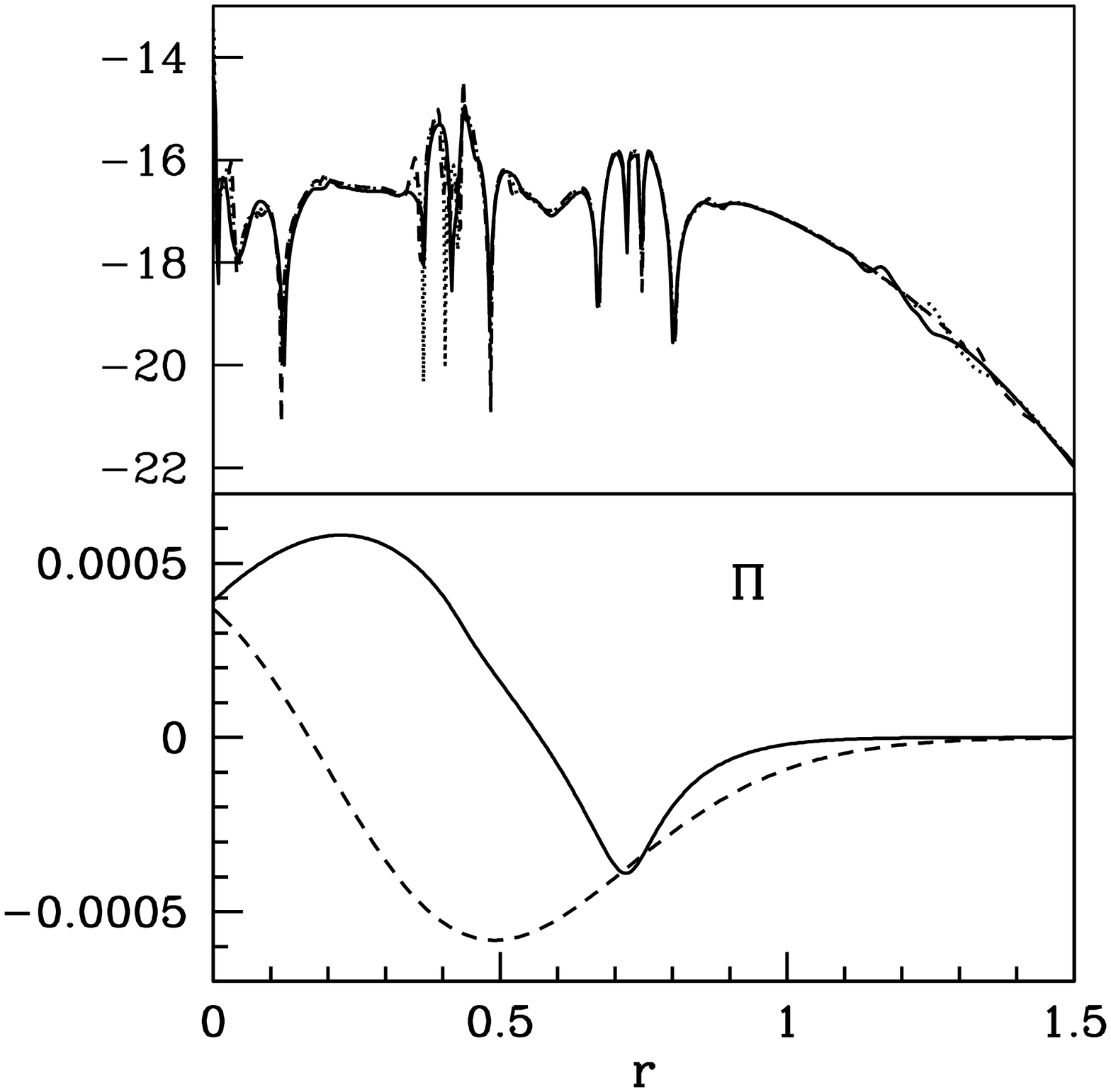}
\caption{Convergence test for the fluid variable $\Pi$.  The top panel 
shows the scaled error estimates described in the caption of Fig.~\ref{fig:D-converge}.  The bottom
panel shows $\Pi(r,0)$ (dashed) and $\Pi(r,t)$ (solid), where $t$ is the time
at which convergence is tested.
\label{fig:Pi-converge}}
\end{figure}

\begin{figure}[htb]
\includegraphics[scale=0.38]{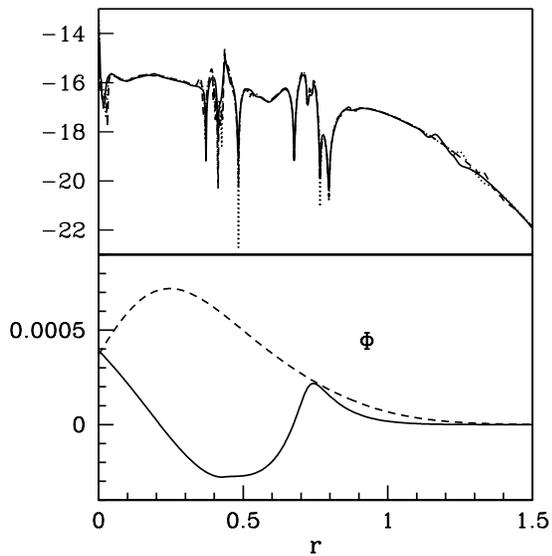}
\caption{Convergence test for the fluid variable $\Phi$.  The top panel 
shows the scaled error estimates described in the caption of Fig.~\ref{fig:D-converge}.  The bottom
panel shows $\Phi(r,0)$ (dashed) and $\Phi(r,t)$ (solid), where $t$ is the time
at which convergence is tested.
\label{fig:Phi-converge}}
\end{figure}

To illustrate convergence properties of our discrete approximations, we show the
results of a convergence test of our code in 
Figs.~\ref{fig:D-converge}--\ref{fig:Phi-converge}, for the quantities
$D$, $\Pi$, and $\Phi$ (\ref{ideal-piphi-state-vectors}), respectively.
The scaled error estimates shown in the 
top panels of each figure demonstrate how the computed value of each dynamical 
variable
exhibits the expected dependence on the fundamental 
grid spacing, $h$.  Specifically, so long as the flow is smooth (including the initial 
conditions), and our spatial and temporal grid spacings are characterized {\em locally}
by a single
discretization scale, $h$, then because our scheme is second order, we expect 
Richardson expansions of the form:
\begin{equation}
	\lim_{h\to 0} f^h(t,r) = f(t,r) + h^2 \, e_2(t,r) + \, \cdots
	\label{richardson}
\end{equation}
where $f(t,r)$ represents any dynamical fluid variable, $f^h(t,r)$ is the 
discrete approximation to that variable, and $e_2(t,r)$ is an 
$h$-independent function with smoothness comparable to $f$. 
The data shown in the plot have been extracted at a time 
before any discontinuities were observed in the solution
to ensure that the assumed Richardson expansions would remain valid.  
In addition, in order to test code convergence in the context of 
the regridding procedure described above, we performed the convergence test at a 
time following the first grid refinement.  From these results, it is evident
that our numerical methods are 2nd-order accurate for smooth flows.  

%%%%%%%%%%%%%%%%%%%%%%%%%%%%%%%%%%%%%%%%%%%%%%%%%%%%%%%%%%%%%%%%%%%%
%%%%%%%%%%%%%%%%%%%%%%%%%%%%%%%%%%%%%%%%%%%%%%%%%%%%%%%%%%%%%%%%%%%%
\section{Pseudo-Code for the Primitive Variable Calculation}
\label{app:pseudo-code-prim}

The primitive variables are calculated from the conserved variables 
using a one-dimensional Newton-Raphson method  
that locates the value of $H$ that minimizes the residual given in 
Eq.~(\ref{pvar-resid2}).  The equations for calculating the primitive variables and other quantities
as a function of the value of $H$ in a given iterations 
are listed in Table~\ref{table:rel-prim-vars-detail} 
for different limits of $\Lambda$. 

\begin{table}
\caption{Pseudo-code for the calculation of 
$\left\{\mathcal{I} , \mathcal{I}^{\,\prime} , P , v , \rho_\circ \right\}$. 
Here, $\mathtt{TAYLOR}_8$ represents the operation of taking 
the series expansion of its argument to $O(b^8)$, which we perform using \texttt{Maple}. Here, $\mathcal{G}\equiv \Gamma/\left(\Gamma-1\right)$.}
\label{table:rel-prim-vars-detail}
\begin{tabbing}
\hspace*{0.5cm}\= \hspace{0.5cm} \= \kill	
\textbf{\texttt{If }} $\left(|\Lambda| > \Lambda_\mathrm{High} \right) 	\ 
                     \mathbf{\mathtt{then}}$ \\[0.25cm]
              \> $b = \frac{1}{ 2 \left|\Lambda\right| } \ , \ B \equiv b^2  $\\
              \> $X(b) \equiv 1 / W^2 = 2 \sqrt{B} \left(\sqrt{B+1} - \sqrt{B}\right)$\\
              \> $\rho_\circ = \mathtt{TAYLOR}_8\left[\frac{D}{a}\sqrt{X(b)}  \right] $\\
              \> $P = \mathtt{TAYLOR}_8\left[ \frac{1}{\mathcal{G}} \left(H - \frac{D}{a}\sqrt{X(b)}\right)  \right] $\\
	      \> $v = \mathtt{sign}(S) \ \mathtt{TAYLOR}_8\left[ \left( \sqrt{B+1} - \sqrt{B} \right) \right] $\\
	      \> $\mathcal{I} = \mathtt{TAYLOR}_8 \left[ H  \left(\frac{1}{X(b)} - \frac{1}{\mathcal{G}}\right) - \tau + D \left( \frac{\sqrt{X(b)}}{a \mathcal{G}} - 1 \right)  \right] $\\
	      \> $\mathcal{I}^\prime = \mathtt{TAYLOR}_8 \left[ \frac{1}{2 \sqrt{B+1}\left(\sqrt{B+1} - \sqrt{B}\right)} - \frac{1}{\mathcal{G}} \right. $ \\ 
              \> $ \left. \quad \quad + \left(\frac{D}{a H \mathcal{G}}\right) \frac{ B^{1/4} \left(\sqrt{B+1} - \sqrt{B}\right)^{3/2}}{\sqrt{2} \sqrt{B+1}} \right] $\\
\textbf{\texttt{Else }}  \\[0.25cm]
\> $\mathbf{\mathtt{If }} \left(  |\Lambda| > \Lambda_\mathrm{Low}\right) 	\ \mathbf{\mathtt{then}}$ \\[0.25cm]
\>\> $\mathcal{Y} \ = \ \sqrt{ 1 + 4 \Lambda^2 } $\\ 
\>\> $v \ = \ \frac{1}{2 \Lambda} \left( \mathcal{Y} - 1 \right) $\\
\>\>  $\pderiv{v}{H} \ = \ - \frac{ S }{ H^2 } \left[ \frac{2}{\mathcal{Y}} - \frac{\left( \mathcal{Y} - 1 \right)}{ 2 \Lambda^2 } \right] $\\[0.25cm]
\>  $\mathbf{\mathtt{Else }}$\\
\>\>  $v \ = \  \left(1+\left(-1+\left(2+\left(-5+\left(14   \right. \right. \right. \right. \right.$\\
\>\>  $ \left. \left. \left. \left. \left.  \quad \quad \quad -42\,{\Lambda}^{2}\right){\Lambda}^{2}\right){\Lambda}^{2}\right){\Lambda}^{2}\right){\Lambda}^{2}\right)\Lambda  $\\[0.25cm]
\>\> $\pderiv{v}{H} \ =  \ - \frac{ S }{ H^2 }  \ \left[ 1+\left(-3+\left(10+\left (-35+\left (126 \right. \right. \right. \right. \right.$ \\ 
\>\> $\left. \left. \left. \left. \left. \quad \quad \quad -462\,{\Lambda}^{2}\right){\Lambda}^{2}\right){\Lambda}^{2}\right){\Lambda}^{2}\right){\Lambda}^{2} \right]  $\\[0.25cm]
\>  $\mathbf{\mathtt{End}} \, \mathbf{\mathtt{If}}$\\[0.25cm]
\>  $W = 1 \, / \sqrt{1 - v^2} \ , \quad P = \frac{1}{\mathcal{G}} \left( H - \frac{ D }{ a W } \right) \ , \quad \rho_\circ = \frac{D}{a W} $\\[0.25cm]
\> $\mathcal{I}(H) = H W^2  \, - \tau - D - P  $ \\[0.25cm]
\>  $\mathcal{I}^{\,\prime}(H) \ = \  W^2 \left( 1 \, + \, 2 H W^2 \, v \, \pderiv{v}{H} \right) \, - \, \frac{1}{\mathcal{G}} \left( 1 \, + \, \frac{D W v}{ a } \, \pderiv{v}{H} \right)  $\\
$\mathbf{\mathtt{End}} \, \mathbf{\mathtt{If}}$
\end{tabbing}
\end{table}

%%%%%%%%%%%%%%%%%%%%%%%%%%%%%%%%%%%%%%%%%%%%%%%%%%%%%%%%%%%%%%%%%%%%%%%%%%%%%%%%%%%%%
%%%%%%%%%%%%%%%%%%%%%%%%%%%%%%%%%%%%%%%%%%%%%%%%%%%%%%%%%%%%%%%%%%%%%%%%%%%%%%%%%%%%%
%%%%%%%%%%%%%%%%%%%%%%%%%%%%%%%%%%%%%%%%%%%%%%%%%%%%%%%%%%%%%%%%%%%%%%%%%%%%%%%%%%%%%
%%%%%%%%%%%%%%%%%%%%%%%%%%%%%%%%%%%%%%%%%%%%%%%%%%%%%%%%%%%%%%%%%%%%%%%%%%%%%%%%%%%%%
%%%%%%%%%%%%%%%%%%%%%%%%%%%%%%%%%%%%%%%%%%%%%%%%%%%%%%%%%%%%%%%%%%%%%%%%%%%%%%%%%%%%%
%%%%%%%%%%%%%%%%%%%%%%%%%%%%%%%%%%%%%%%%%%%%%%%%%%%%%%%%%%%%%%%%%%%%%%%%%%%%%%%%%%%%%
%% BIBLIOGRAPHY:   %%%%%%%%%%%%%%%%%%%%%%%%%%%%%%%%%%%%%%%%%%%%%%%%%%%%%%%%%%%%%%%
%%%%%%%%%%%%%%%%%%%%%%%%%%%%%%%%%%%%%%%%%%%%%%%%%%%%%%%%%%%%%%%%%%%%%%%%%%%%%%%%%%%%%
%%%%%%%%%%%%%%%%%%%%%%%%%%%%%%%%%%%%%%%%%%%%%%%%%%%%%%%%%%%%%%%%%%%%%%
%                                                                    %
% NOTE: For master's theses and reports, NOTHING is permitted to     %
%       come between the bibliography and the vita. The command      %
%       to generate the index (if used) MUST be moved to before      %
%       this section.                                                %
%                                                                    %
%\nocite{*}      % This command causes all items in the               %
                % bibliographic database to be added to              %
                % the bibliography, even if they are not             %
                % explicitly cited in the text.                      %
                %                                                    %
%\index{Bibliography@\emph{Bibliography}}%                            %


\begin{thebibliography}{999}
%%%%%%%%%%%%%%%%%%%%%%%%%%%%%%%%%%%%%%%%%%%%%%%%%%%%%%%%%%%%%%%%%%%%%%

%CP review - intro., theory
%Matt's review of crit. phen. 
\bibitem{choptuik-1998} M.~W.~Choptuik, 
%``The (Unstable) Threshold of Black Hole Formation,''
\grqc{9803075}, (1998).

\bibitem{gundlach} C.~Gundlach, 
%        ``Critical Phenomena in Gravitational Collapse,'' \\
\lr, \url{http://www.livingreviews.org/lrr-1999-4} (1999);
\grqc{0210101} (2002).

\bibitem{gundlach-rev2} C.~Gundlach, 
%``Critical Phenomena in Gravitational Collapse,''
	\physrep, \textbf{376}, 339-405 (2002)
%        \url{http://arxiv.org/ps/gr-qc/0210101} 

%CP - intro., theory
%Matt's original paper on EMKG Crit. phen. 
\bibitem{choptuik-1993} M.~W.~Choptuik, \prl, \textbf{70}, 9 (1993). 

%CP -- intro.
%Tie between and discussion of  crit. phen. of scalar field and \Gamma=2 fluid;
\bibitem[Brady et al.(2002)Brady, Choptuik, Gundlach, and Neilsen]{brady_etal} P.~R.~Brady, 
M.~W.~Choptuik, C.~Gundlach, and D.~W.~Neilsen, 
%``Black-hole Threshold Solutions in Stiff Fluid Collapse,'' 
\cqg, \textbf{19}, 6359-6375 (2002). 
%\url{http://arxiv.org/ps/gr-qc/0207096}, (2002).

\bibitem{evans-coleman} C.~R.~Evans and J.~S.~Coleman, 
%``Critical Phenomena and Self-Similarity in the Gravitational Collapse of Radiation Fluid,''
\prl, \textbf{72}, 1782-1785 (1994). 

\bibitem{neilsen-crit} D.~W.~Neilsen and M.~W.~Choptuik,  
%``Critical Phenomena in Perfect Fluids,''
\cqg, \textbf{17}, 761-782 (2000).

%Yang-mills type-I/II original;
\bibitem{choptuik-chmaj-bizon} M.~W.~Choptuik, T.~Chmaj, and P.~Bizo\'{n}, 
%``Critical Behavior in Gravitational Collapse of a Yang-Mills Field,''
\prl, \textbf{77}, 424-427 (1996). 

%CP -- type-I link to our I/II boundary
%Study the supercritical regime to differentiate between the type-I and type-II behavior in EYM;
% Found that the barrier between the types is a new critical solution called a colored black 
% hole.  It's a black hole solution with YM hair that has a non-trivial profile and are 
% are characterized by the number zero-crossings they have like the Bartnik-Mckinnon solutions. 
\bibitem{choptuik-hirshmann-marsa} M.~W.~Choptuik, E.~W.~Hirschmann, and R.~L.~Marsa, 
%``New Critical Behavior in Einstein-Yang-Mills Collapse,''
\prd, \textbf{60}, 124011 (1999). 

\bibitem{olabarrieta-choptuik} I.~Olabarrieta and M.~W.~Choptuik, 
%``Critical Phenomena at the Threshold of Black Hole Formation for Collisionless Matter in Spherical Symmetry,'' 
\prd, \textbf{65}, 024007 (2001). 

\bibitem{rein-etal-1998} G.~Rein, A.~D.~Rendall, and J.~Schaeffer, 
%``Critical Collapse of Collisionless Matter:  A Numerical Investigation,''
\prd, \textbf{58}, 044007 (1998).


\bibitem{noble-choptuik2} S.~C.~Noble and M.~W.~Choptuik, 
%``Driven Neutron Star Collapse: Type-I Critical Phenomenon to Initial Black Hole Mass Distribution,'' 
\textit{in preparation} (2007). 

\bibitem{koike-etal-1995} T.~Koike, T.~Hara, and S.~Adachi, 
%``Critical Behavior in Gravitational Collapse of Radiation Fluid:  A Renormalization Group (Linear Perturbation) Analysis,'' 
\prl, \textbf{74}, 5170-5173 (1995).

%Continued the analysis (CSS ansatz) that Evans and Coleman used to extend study for 0.01<=k<=0.888;
\bibitem{maison-1996} D.~Maison, 
%``Non-universality of Critical Behavior in Spherically Symmetric Gravitational Collapse,''
\physlettb, \textbf{366}, 82-84 (1996). 

%Mentions in Discussion a little about Scalar+Fluid and weak scalar allowing for crit. sol. to be stable.
\bibitem{hara-etal-1996}  T.~Hara, T.~Koike, and S.~Adachi, 
%``Renormalization Group and Critical Behavior in Gravitational Collapse,''
\grqc{9607010} (1996). 

%discusses all \gamma for all \Gamma;
% Also, give general EOS that would yield same universality class. 
\bibitem{koike-etal-1999} T.~Koike, T.~Hara, and S.~Adachi, 
%``Critical Behavior in Gravitational Collapse of Perfect Fluid,''
\prd, \textbf{59}, 104008 (1999).

%CP -- intro., discussion of CSS coordinates;
%allows for a self-similar (of first kind) solution.  
%Show that barotropic P=k\rho EOS is the only one to admit self-similar solutions of this kind;
\bibitem{cahill-taub} M.~E.~Cahill and A.~H.~Taub, 
\cmp, \textbf{21}, 1 (1971).

%First example of self-similar collapse given explicit scale-dependence and 
\bibitem{choptuik-1994} M.~W.~Choptuik, 
in \textit{Deterministic Chaos in General Relativity}, 
edited by D.~Hobill et al., 155 (1994). 

\bibitem{novak} J.~Novak, 
%``Velocity-induced collapses of stable neutron stars,'' 
\aap, \textbf{276}, 606-613 (2001).
% \url{http://arxiv.org/ps/gr-qc/0107045}.

%Description of realistic, tabulated EOS used by Novak
\bibitem{pons-etal-2000} J.~A.~Pons, S.~Reddy, P.~J.~Ellis, M.~Prakash, and J.~M.~Lattimer, 
%``Kaon Condensation in Proto-Neutron Star Matter,''
\prc, \textbf{62}, (2000). 

\bibitem{wald} R.~M.~Wald, \textit{General Relativity} 
(University of Chicago Press, Chicago, 1984).

%GR CFD methods:
%Valencia formulation of local characteristic equations for Hydro equations in 3+1 formalism 
% in flux-conservative form.  And various tests: Sod, spherical  and nonspherical dust accretion, 
\bibitem[Banyuls et al.(1997)Banyuls, Font, Ib\'{a}\~{n}ez, and Mart\'{i}]{banyuls} 
F.~Banyuls,  J.~A.~Font, \ibanez,  \marti, and
J.~A.~Miralles, \apj, \textbf{476}, 221-231, (1997).

%GR CFD -- review 
\bibitem{font-review} J.~A.~Font, 
%``Numerical Hydrodynamics in General Relativity,'' \\
\lr, \url{http://www.livingreviews.org/lrr-2003-4}
(2003).

%3D Type~I, 3D GR CFD
%Compares and contrasts Marquina to Roe, ideal-gas to adiabatic EOS's, for long-term 
%stellar evolutions, collapses of unstable stars to stable ones, rapidly rotating stars and 
% their quasi-radial modes.  Obtained for the first time the eigenfrequencies  of rotating 
% stars in GR and rapid rotation. 
\bibitem{font-etal2} J.~A.~Font, T.~Goodale, S.~Iyer, M.~Miller, L.~Rezzolla, E.~Seidel, 
N.~Stergioulas, W.~Suen, and M.~Tobias, 
\prd, \textbf{65}, 084024 (2002). 

\bibitem{neilsen} D.~W.~Neilsen and M.~W.~Choptuik,  
%``Ultrarelativistic fluid dynamics,''
\cqg, \textbf{17}, 733-759 (2000).

\bibitem{romero} J.~V.~Romero, \ibanez, \marti, and J.~A.~Miralles, 
\apj, \textbf{462}, 839-854 (1996).

\bibitem[Arnowitt et al.(1962)Arnowitt, Deser, and Misner]{adm} R.~Arnowitt, S.~Deser, and C.~W.~Misner, 
%``The Dynamics of General Relativity,'' 
in \textit{Gravitation: an Introduction to Current Research}, 
edited L. Witten (John Wiley \& Sons, New York, 1962). 

%TOV paper#1:
\bibitem{oppenheimer-volkoff} J.~R.~Oppenheimer and G.~M.~Volkoff, 
%``On Massive Neutron Cores,''  
\pr, \textbf{55}, 374-381 (1939). 

% TOV paper#2:
\bibitem{tolman-book} R.~C.~Tolman, 
\textit{Relativity, Thermodynamics and Cosmology} (Oxford University Press, 1934). 

% TOV paper#2:
\bibitem{tolman-paper} R.~C.~Tolman, 
%``Static Solutions of Einstein's Field Equations for Spheres of Fluid,''
\pr, \textbf{55}, 364-373 (1939).

%Rotating case of TOV solutions:
%Gives scaling with respect to K of the EOS
% By using equilibrium models of uniformly and nonuniformly rotating axi-symmetric stars. 
% Stable Spin-up from angular mom. loss (the star widens) can be seen close to Gamma=4/3 
% and larger values for differentially rotating stars.  Also, more massive stars can 
% lose ang. mom., spin-up and mass-shed before rot. speed is high enough to undergo quasi-radial
% instability and collapse.  
\bibitem{cook-shap-teuk-1992} G.~B.~Cook, S.~L.~Shapiro, and S.~A.~Teukolsky,
\apj, \textbf{398}, 203-223 (1992).

%Uses pseudo-spectral code, used tabulated EOS's (see gourg1), 
% -- used in-going coordinate velocity in order to study Shap&Teuk's question;
% -- found 1)small bounce/oscillation, 2)shock/bounce/? (couldn't resolve shock)  3)prompt coll.
% -- found that M < M_max stars could form black holes;
\bibitem{gourg2} E.~Gourgoulhon, 
%``1D Numerical Relativity applied to neutron star collapse,'' 
\cqg, \textbf{9}, 117-125 (1992).

\bibitem{noble} S.~C.~Noble, 
%``A Numerical Study of Relativistic Fluid Collapse,'' 
Ph.D. thesis, The University of Texas at Austin, 2003;
\grqc{0310116} (2003).

\bibitem{marsa-choptuik} R.~L.~Marsa and M.~W.~Choptuik, 
\textit{The RNPL Reference Manual}, 
\url{http://bh0.physics.ubc.ca/People/marsa/rnpl/refman/refman.html} (1995); 
\textit{The RNPL User's Guide}, 
\url{http://bh0.physics.ubc.ca/People/marsa/rnpl/users_guide/users_guide.html} (1995). 

\bibitem{roe-1981} P.~L.~Roe, 
%``Approximate Riemann Solvers, Parameter Vectors, and Difference  Schemes'', 
\jcompphys, \textbf{43}, 357 (1981).

%CFD
% Marquina method source :
\bibitem{donat-marquina} R.~Donat  and  A.~Marquina, 
\jcompphys, \textbf{125}, 42-58 (1996).

%Entropy fix for the Roe's Approx. Riemann solver;
\bibitem{harten-hyman} A.~Harten and J.~M.~Hyman, 
%``Self-Adjusting Grid Methods for One-Dimensional Hyperbolic Conservation Laws,'' 
\jcompphys, \textbf{50}, 235-269 (1983).

% minmod limiter: 
\bibitem{vanleer-1979} B.~van~Leer
\jcompphys, \textbf{32}, 101-136 (1979).

% MC limiter: 
\bibitem{vanleer-1977} B.~van~Leer
\jcompphys, \textbf{23}, 276-299 (1977).

%Superbee limiter:
\bibitem{roe-1985} P.~L.~Roe, 
Lect. Notes Appl. Math., \textbf{22}, 163-193 (1985).

\bibitem{shu1997} C.-W.~Shu, 
%``Essentially Non-Oscillatory and Weighted Essentially Non-Oscillatory Schemes forHyperbolic Conservation Laws'', 
NASA CR-97-206253 ICASE Report No. 97-65, p. 83 (1997);
\url{ftp://ftp.icase.edu/pub/techreports/97/97-65.ps}.

\bibitem{neilsen-thesis} D.~W.~Neilsen, 
%``Extremely Relativistic Fluids in Strong-Field Gravity,''
Ph.D. thesis, The University of Texas at Austin, 1999.

\bibitem{olabarrieta} I.~Olabarrieta, ENO computer code, 
\url{http://laplace.physics.ubc.ca/People/inaki/fluids/eno.html} 
(2002).

\bibitem{eulderink} F.~Eulderink  and G.~Mellema, \aaps, \textbf{110}, 587 (1995). 

\bibitem{leveque1} R.~J.~LeVeque, 
\textit{Numerical Methods for Conservation Laws} 
(Birkha\"{u}ser-Verlag, Basel, 1992).

\bibitem{leveque2} R.~J.~LeVeque, in \textit{Computational Methods 
for Astrophysical Fluid Flow: 27th Saas-Fee Advanced Course Lecture Notes}
(Springer-Verlag, Berlin, 1998).

\bibitem{quirk} J.~J.~Quirk, 
%``A contribution to the great Riemann solver debate,'' 
\ijnmf, \textbf{18}, 555-574 (1994).

%GR CFD methods:
% Demonstrate differences between Roe and Marq. 
\bibitem{donat-font} R.~Donat, J.~A.~Font, \ibanez, and A.~Marquina, 
\jcompphys, \textbf{146}, 58-81 (1998).

\bibitem{marti-muller-living} 
\marti, E.~M\"{u}ller, 
%``Numerical Hydrodynamics in Special Relativity,''
\lr, \url{http://www.livingreviews.org/lrr-2003-7} (2003).

%First instance of Eulerian-type code in curved spacetimes; studied accretion onto a stationary 
% Kerr hole in axisymmetry;
\bibitem{wilson2} J.~R.~Wilson, 
%``Numerical Study of Fluid Flow in a Kerr Space,''
\apj, \textbf{173}, 431-438 (1972).

%CP, intro.
%Conjecture and show for EMKG that curvature scalar R scales as |p-p*|^(-2\gamma);
\bibitem{garfinkle2} D.~Garfinkle and G.~C.~Duncan, 
%``Scaling of curvature in subcritical gravitational collapse,'' 
\prd, \textbf{58}, 064024 (1998). 

%Kink instability for \Gamma>0.89:
\bibitem{harada-2001} T.~Harada, 
%``Stability Criterion for Self-Similar Solutions with Perfect Fluids in General Relativity,''
\cqg, \textbf{18}, 4549-4567 (2001).

%\bibitem{liebling-choptuik-1996} S.~L.~Liebling and M.~W.~Choptuik, 
%%``Black Hole Criticality in the Brans-Dicke Model,''
%\prl, \textbf{77}, 1424 (1996).

%\bibitem{hirschmann-eardley-1997} E.~W.~Hirschmann and D.~M.~Eardley, 
%%``Criticality and Bifurcation in the Gravitational Collapse of a Self-Coupled Scalar Field,''
%\prd, \textbf{56}, 4696 (1997).

\bibitem{liebling-1998} S.~L.~Liebling, 
%``Multiply Unstable Black Hole Critical Solutions,''
\prd, \textbf{58}, 084015 (1998).

%Graxi
\bibitem{choptuik-etal-2003} M.~W. Choptuik, E.~W. Hirschmann, S.~L. Liebling, and F.~Pretorius, 
%``An Axisymmetric Gravitational Collapse Code,'' 
\cqg, \textbf{20}, 1857-1878 (2003).
%\url{http://arxiv.org/ps/gr-qc/0301006} (2003). 


\end{thebibliography}
\end{document}